\documentclass[preprints,article,accept,moreauthors,pdftex]{mdpi}%{Definitions/mdpi}%,,submit,
\usepackage{amsmath}
\usepackage{amsfonts}
\usepackage{amssymb,bm}
\usepackage{siunitx}
\usepackage{color}
\usepackage{braket}
\usepackage{graphics}
\UseRawInputEncoding
\firstpage{1}
\pubvolume{8}
\issuenum{11}
\articlenumber{574}
\pubyear{2022}
\copyrightyear{2022}
\datereceived{2 October 2022}
\dateaccepted{27 October 2022}
\datepublished{31 October 2022}
\hreflink{https://doi.org/10.3390/universe8110574} % If needed use \linebreak
\Title{{\bf\textit {Experimentum crucis}} for electromagnetic response of metals to
evanescent waves and the Casimir puzzle}

\TitleCitation{{\textit {Experimentum crucis}} for electromagnetic response of metals to
evanescent waves and the Casimir puzzle}

\Author{
Galina L. Klimchitskaya ${}^{1,2}$,
Vladimir M. Mostepanenko ${}^{1,2,3}$,
Vitaly B. Svetovoy ${}^4$}

\AuthorNames{
Galina L. Klimchitskaya, Vladimir M. Mostepanenko, Vitaly B. Svetovoy}

\AuthorCitation{
Klimchitskaya, G.L.; Mostepanenko, V.M.; Svetovoy, V.B.}

\address{%
${}^{1}$\quad Central Astronomical Observatory at Pulkovo of the Russian Academy of Sciences, Saint Petersburg, 196140, Russia\\
${}^{2}$\quad Peter the Great Saint Petersburg
Polytechnic University, Saint Petersburg, 195251, Russia\\
${}^{3}$\quad Kazan Federal University, Kazan, 420008, Russia\\
${}^4$\quad Frumkin Institute of Physical Chemistry and Electrochemistry,
Russian Academy
of Sciences, Leninsky prospect 31 bld. 4, 119071 Moscow, Russia}

\corres{Correspondence: vmostepa@gmail.com}

\abstract{It is well known that the Casimir force calculated at large separations
using the Lifshitz theory differs by a factor of 2 for metals described
by the Drude or plasma models. We argue that this difference
is entirely determined by the contribution of transverse electric
($s$) evanescent waves. Taking into account that there is a lack
of experimental information on the electromagnetic response
of metals to low-frequency evanescent waves, we propose an
experiment on measuring the magnetic field of an oscillating
magnetic dipole spaced in vacuum above a thick metallic plate.
According to our results, the lateral components of this field are
governed by the transverse electric evanescent waves and may vary
by orders of magnitude depending on the model describing the
permittivity of the plates used in calculations and the
oscillation frequency of the magnetic dipole.
Measuring the lateral component of the magnetic field for
typical parameters of the magnetic dipole designed in the form
of 1-mm coil, one could either validate or disprove
applicability of the Drude model as a response function of
metal in the range of low-frequency evanescent waves. This
will elucidate the roots of the Casimir puzzle lying in the
fact that the theoretical predictions of the Lifshitz theory
using the Drude model are in contradiction with the
high-precision measurements of the Casimir force at
separations exceeding 150~nm. Possible
implications of the suggested experiment for a wide range
of topics in optics and condensed matter physics dealing with
evanescent waves are discussed.}


\begin{document}
%%%%%%%%%%%%%%%%%%%%%%%%%%%%%%%%%%%%%%%%%%
\section{Introduction}

Casimir's discovery \cite{1} that two parallel uncharged ideal-metal planes
attract each other marked the beginning of active research and led to new
concepts and far-reaching implications in both fundamental and applied physics
(see, e.g., the monographs \cite{2,3,4,5}). The Casimir force results from
zero-point and thermal (if the planes are kept at nonzero temperature)
fluctuations of the electromagnetic field. On ideal-metal planes the
tangential component of electric field and the normal component of
magnetic induction vanish leading to the discrete oscillation frequencies.
A finite difference between the infinite free energies in the presence and
in the absence of the planes determines the finite Casimir free energy per
unit area. The derivative of this free energy with respect to the
separation between the planes with a minus sign is the Casimir force.

The Lifshitz theory generalized the Casimir force for two thick material
plates (semispaces) in thermal equilibrium with the environment \cite{6,6a,7}.
In this case, the electromagnetic response of the plate materials is given
by the standard continuity boundary conditions formulated in terms of
the frequency-dependent dielectric permittivity and (for magnetic plates)
magnetic permeability. The Lifshitz theory treats the van der Waals
force as a particular case of the Casimir force at separations of a few
nanometers where one can neglect the effects of relativistic
retardation. During the last two decades the Lifshitz theory was
generalized for material bodies of arbitrary shape \cite{8,9,10}.

The interest in measuring the Casimir force was rekindled by the
experiment of   \cite{11a} (an unaccounted systematic error which
was present in this experiment was corrected by its author in
  \cite{11b}). The beginning of the 21st century was marked by performing
several high-precision experiments on measuring the Casimir force between
metallic test bodies. These experiments were made by means of a
micromechanical torsional oscillator \cite{11,12,13,14,15,16} and
an atomic force microscope \cite{17,18,19,20,21,22,23} at the
separation distances exceeding 150~nm. A comparison
between the measurement data and theoretical predictions of the
Lifshitz theory showed that these predictions are experimentally
excluded if the electromagnetic response of metals (Au, Ni) at low
frequencies is described by the well-established Drude model which
takes the proper account of the relaxation properties of conduction
electrons \cite{11,12,13,14,15,16,17,18,19,20,21,22,23}. If,
however, the response of metals at low frequencies is described by
the dissipationless plasma model, a very good agreement between
the measurement data and theoretical predictions was revealed
\cite{11,12,13,14,15,16,17,18,19,20,21,22,23}.

For distances below 100~nm, a comparison between the Lifshitz
theory and the measurement data demonstrated good agreement
(see, e.g., \cite{23a,23b,23c,23d}). Analysis made it
apparent that the problem is related to the low-frequency
contribution to the Casimir force, which is determined by
conduction electrons \cite{57,59,63}.
At separations below 100~nm the major
contribution to the Casimir force is given by the core
electrons whereas with increasing separation the role of
conduction electrons progressively increases. Thus, it is very
surprising that the Drude model does not work at large
separations because it provides a well-established description
of collisions of conduction electrons which are responsible
for the resistance of metals (Ohm's law).

No less surprise is the fact that for metals with perfect
crystal lattices described by the Drude model the Casimir entropy
goes with vanishing temperature to the negative quantity which
depends on the parameters of a system, such as the separation
between the plates and the plasma frequency, i.e., the Nernst
heat theorem is violated \cite{24,25,26,27}. If the
experimentally consistent plasma model is used, the Nernst heat
theorem is satisfied \cite{24,25,26,27} in spite of the fact that
this model disregards the dissipation of conduction electrons, which
plays an important role at low frequencies. To satisfy
the Nernst heat theorem with the Drude model, it was suggested
to take into account a nonzero residual relaxation at zero
temperature, which exists for crystal lattices with defects
\cite{28,29,30}. It was noted, however, that the
perfect crystal lattice is an equilibrium system with the
nondegenerate ground state so that the Nernst heat theorem
must be satisfied in this case \cite{31}. Moreover, in
equilibrium there must be no defects \cite{31a}.
The unresolved problems of disagreement between the
theoretical predictions of the Lifshitz theory and
measurement data, on the one hand, and the requirements of
thermodynamics, on the other hand, were named the Casimir puzzle
\cite{32,33,33a}.

During the last 20 years, many attempts to resolve the Casimir
puzzle have been undertaken based on the assumptions that there
might be some unaccounted systematic effects in the performed
experiments or inaccuracies in the comparison between the
measurement data and theoretical predictions
(see   \cite{4,34,35,36} for review).
These efforts, although did not
furnish the resolution of the Casimir puzzle, have had a
profound impact on future investigations of the problem.
Specifically, it was underlined that in the frequency region of
the anomalous skin effect the concept of the dielectric
permittivity depending only on frequency used in the standard
Lifshitz theory loses its meaning and one has to take into
account the effects of spatial dispersion
\cite{31a,37,38,39,40,41,42,43}. In doing so, the reflection
coefficients entering the Lifshitz formula for the Casimir
pressure are expressed via the nonlocal surface impedances
\cite{37,40}. This approach allowed some progress in resolution
of thermodynamic part of the Casimir puzzle \cite{41,42} but
did not remove the contradiction between experiment and theory
because for metallic test bodies corrections to the Casimir pressure due to spatial
nonlocality in the region of the anomalous skin effect turned out to be too
small \cite{39,40}.

To find a solution for both components of the Casimir puzzle,
recently the phenomenological spatially nonlocal dielectric
permittivities have
been proposed \cite{44,45,46} which almost coincide with the
permittivity of the standard Drude model for the propagating
waves satisfying the condition $k_{\bot}<\omega/c$ where
$k_{\bot}$ is the magnitude of the wave vector projection on
the plane of Casimir plates and $\omega$ is the frequency. The
suggested permittivities, however, can deviate from the Drude
model significantly for the evanescent waves which obey the
inequality $k_{\bot}>\omega/c$ and, thus, are
characterized by a pure imaginary $k_z$. Allowing for only
real $k_z$, one can say that the evanescent waves are off the
mass shell in the free space. It was shown that the Lifshitz
theory using the permittivities of this kind is in
agreement with the Nernst heat theorem \cite{47} and with the
measurement data of all high-precision experiments on
measuring the Casimir force \cite{44,45,46}.
Thus, we can say that the nonlocal permittivities, proposed for
metals in \cite{44,45,46}, are to some extent analogous to the
permittivities of graphene, which are also spatially nonlocal
but were derived on the basis of first principles of quantum
electrodynamics at nonzero temperature using the formalism
of polarization tensor \cite{81,82}. It was shown that for graphene
described by these permittivities there is no Casimir puzzle, i.e.,
the Lifshitz theory is consistent with both the experimental results
\cite{84,85} and the requirements of thermodynamics \cite{83,83a}.

The proposed phenomenological
permittivities \cite{44,45,46} take into account the dissipation of conduction
electrons in metals and leave almost unchanged the transverse
magnetic ($p$) reflection coefficient as compared to the
standard Drude model. As to the transverse electric ($s$)
reflection coefficient, its value at zero frequency becomes
nonzero for both propagating and evanescent waves as it holds
if the plasma model is used (recall that for evanescent waves the
transverse electric reflection coefficient calculated using the
Drude model vanishes at zero frequency). Keeping in mind that the
suggested nonlocal permittivities are of entirely phenomenological
character, it is desirable to determine their possible role in
some physical phenomena other than the Casimir effect.

The direct measurement of the $s$-type reflection coefficient in the
range of evanescent waves cannot be performed because all conventional
methods (for example, ellipsometry) are designed for propagating
waves. Nevertheless,
the evanescent fields are actively discussed in the literature
in relation to the possibility to overcome the diffraction
limit in optics \cite{52}. Although physics of surface
plasmon polaritons \cite{48} provides a considerable body of
data related to the region of rather large $k_{\bot}$, it is
restricted to only the transverse magnetic ($p$) reflection
coefficient. The widely used techniques of total internal
reflection and frustrated total internal reflection give the
possibility to test the reflection properties in the region of
$k_{\bot}$ only slightly exceeding $\omega/c$. This is
connected with the fact that there are no transparent media
with sufficiently large index of refraction (see the
long-performed experiments \cite{49,50,51}). The method of
nano frustrated total internal reflection exploits the
illumination of a nanoparticle of size $R \sim 1/k_{\bot}$
placed near the material surface under investigation. This
approach is employed in the near-field optical microscopy,
which allows to exceed the standard resolution limit using
the evanescent waves \cite{52,53}. The
near-field optical microscopes, however, are mostly
employed in various technological applications \cite{53}
and are more sensitive to the $p$-polarized electromagnetic
field \cite{54}.

In this paper, we suggest the {\it experimentum crucis}, which can
reliably select between different response functions of metals
to the low-frequency electromagnetic field in the area of
evanescent waves. To do so, we consider an oscillating
magnetic dipole spaced at a distance of a few millimeters from
a metallic plate. The electric field of this dipole is
negligibly small as compared to the magnetic one. We obtain
explicit expressions for the latter via the transverse electric
($s$) reflection coefficient of the plate without specifying its
form. According to our results, the lateral field components are fully
determined by the contribution of extremely evanescent waves
$(k_{\bot}\gg\omega/c)$, whereas the
contribution of propagating waves is negligibly
small. The components of the magnetic field are calculated
as functions of lateral separation from the dipole
for different oscillation frequencies using the
dielectric permittivities of metal described by the Drude
model, plasma model, and the spatially nonlocal model of
  \cite{44,45,46,47}. It is shown that the lateral field
components, which are fully determined by reflections on the
plate, strongly depend on the model of used dielectric
permittivity. For example, the lateral field component computed
using the Drude model is smaller by up to a factor of $10^4$
than that computed using the plasma model or the spatially
nonlocal model of   \cite{44,45,46,47}. Thus, by
measuring the field of oscillating magnetic dipole spaced near
a metallic plate, one can probe the response function of metal
in the range of evanescent waves.

The above results are used to propose the {\it experimentum crucis}
for the Casimir puzzle. For this purpose, we consider
the Casimir pressure at large separations between the plates.
It is demonstrated that the differences in theoretical predictions
obtained using the Drude model, plasma model and the spatially
nonlocal model of   \cite{44,45,46,47} are fully determined
by different contributions from the evanescent waves to the
transverse electric ($s$) reflection coefficient. Possible
parameters of the experiment, which could determine the
response function of metals to the extremely evanescent waves, are
suggested. The role of this experiment not only for the
Casimir physics, but for a wide range of topics in optics and
condensed matter physics is discussed.

The brief presentation of the part of obtained results related to
the lateral component of the field of magnetic dipole is published
in the form of a letter in   \cite{letter}.

The paper is organized as follows. In Section~2, it is shown that
the conflict between experiment and theory in the Casimir physics
is determined by a contribution from the evanescent waves.
Section~3 contains the derivation of the field of oscillating
magnetic dipole spaced near a metallic plate. In Section~4, we
compute the dipole field for different frequencies and find how
the contribution from the evanescent waves depends on the
response function of a metal. Section~5  suggests the
{\it experimentum crucis} for the Casimir puzzle.
 In Sections~6 and 7,
the reader will find  a discussion and our conclusions. The
Appendices A and B contain some  necessary details
of the mathematical derivations.

%%%%%%%%%%%%%%%%%%%%%%%%%%%%%%%%%%%%%%%%%%%%%%%%%%%%%%%%%%%%%%%%%%%%%%%%%%
\newcommand{\kb}{{k_{\bot}}}
\newcommand{\skb}{{k_{\bot}^2}}
\newcommand{\vk}{{\mbox{\boldmath$k$}}}
\newcommand{\rv}{{\mbox{\boldmath$r$}}}
\newcommand{\ve}{{\varepsilon}}
\newcommand{\okb}{{(\omega,k_{\bot})}}
\newcommand{\dokb}{{(\omega_d,k_{\bot})}}

%%%%%%%%%%%%%%%%%%%%%%%%%%%%%%%%%%%%%%%%%%%%%%%%%%%%%%%%%%%%%%%%%%%%%%%%%%
\section{Role of evanescent waves in the conflict between experiment and
theory in Casimir physics}
It has been known that the Casimir pressure between two thick material plates
(semispaces) maintained at separation $a$ at temperature $T$ in thermal
equilibrium with the environment is given by the Lifshitz formula which can
be written in two mathematically equivalent representations.
The representation that has enjoyed the widest application is in terms of the
pure imaginary discrete Matsubara frequencies
$\omega_l=i\xi_l=2\pi i k_BTl/\hbar$ with $l=0,\,1,\,2,\,\ldots$
\cite{4,5,6,6a,7}
\begin{equation}
P(a,T)=-\frac{k_BT}{\pi}\sum_{l=0}^{\infty}{\vphantom{\sum}}^{\prime}
\!\!\int_0^{\infty}\!\!\!d\kb q_l\kb
\sum_{\alpha}
\frac{r_{\alpha}^2(i\xi_l,\kb)e^{-2aq_l}}{1-r_{\alpha}^2(i\xi_l,\kb)e^{-2aq_l}},
\label{eq1}
\end{equation}
% \noindent
where the magnitude of the lateral wave vector component
$\kb=(k_x^2+k_y^2)^{1/2}$ was already defined in Section~ 1,
$q_l=(\skb+\xi_l^2/c^2)^{1/2}$,  the prime on the summation sign in $l$
divides the term with $l=0$ by 2, $k_B$ is the Boltzmann constant, and the sum
in $\alpha$ is over the transverse magnetic (TM or $p$) and transverse electric
(TE or $s$) polarizations of the electromagnetic field.
The respective reflection coefficients are given by
\begin{equation}
r_{\rm TM}(i\xi_l,\kb)=\frac{\ve_lq_l-p_l}{\ve_lq_l+p_l},
\qquad
r_{\rm TE}(i\xi_l,\kb)=\frac{q_l-p_l}{q_l+p_l},
\label{eq2}
\end{equation}
% \noindent
where $\ve_l\equiv\ve(i\xi_l)$ is the dielectric permittivity of the plates
calculated at the Matsubara frequencies and
$p_l=p_l(\kb)=(\skb+\ve_l\xi_l^2/c^2)^{1/2}$. The signs of the coefficients (\ref{eq2})
are chosen in such a way that in the limiting case of ideal metal planes
$r_{\rm TM}=1$ and $r_{\rm TE}=-1$.

Another representation of the Lifshitz formula is in terms or real frequencies
\cite{4,5,6,6a}. It expresses the Casimir pressure as the sum of contributions
from the propagating and evanescent waves
\begin{equation}
P(a,T)=P^{\rm prop}(a,T)+P^{\rm evan}(a,T),
\label{eq3}
\end{equation}
\noindent
where
\begin{equation}
P^{\rm prop}(a,T)=-\frac{\hbar}{2\pi^2}\int_{0}^{\infty}\!\!d\omega
\coth\frac{\hbar\omega}{2k_BT}
\int_0^{k_0}\!\!d\kb\kb \sum_{\alpha}{\rm Im}\left[q
\frac{r_{\alpha}^2\okb e^{-2aq}}{1-r_{\alpha}^2\okb e^{-2aq}}\right]
\label{eq4}
\end{equation}
\noindent
and
\begin{equation}
P^{\rm evan}(a,T)=-\frac{\hbar}{2\pi^2}\int_{0}^{\infty}\!\!d\omega
\coth\frac{\hbar\omega}{2k_BT}
\int_{k_0}^{\infty}\!\!d\kb\kb q\sum_{\alpha}{\rm Im}
\frac{r_{\alpha}^2\okb e^{-2aq}}{1-r_{\alpha}^2\okb e^{-2aq}}.
\label{eq5}
\end{equation}
\noindent
Here, $q=q\okb=(\skb-k_0^2)^{1/2}$, $k_0=\omega/c$
and the reflection coefficients (\ref{eq2})
with  $p=p\okb=[\skb-\ve(\omega)k_0^2]^{1/2}$
can be transformed to the familiar Fresnel coefficients \cite{55}
\begin{equation}
r_{\rm TM}\okb=\frac{\ve(\omega)q-p}{\ve(\omega)q+p}=
\frac{\ve(\omega)\tilde{q}-\tilde{p}}{\ve(\omega)\tilde{q}+\tilde{p}},
\qquad
r_{\rm TE}\okb=\frac{q-p}{q+p}=
\frac{\tilde{q}-\tilde{p}}{\tilde{q}+\tilde{p}},
\label{eq6}
\end{equation}
\noindent
where
\begin{equation}
\tilde{q}=\tilde{q}\okb=(k_0^2-\skb)^{1/2},
\qquad
\tilde{p}=\tilde{p}\okb=[\ve(\omega)k_0^2-\skb]^{1/2}.
\label{eq7}
\end{equation}

Note, that $\tilde{q}=iq$ and $\tilde{p}=ip$. Because of this, equation (\ref{eq4}),
expressing the contribution of propagating waves in the Casimir pressure, contains
the rapidly oscillating function $\exp(2ia\tilde{q})$ and is not convenient
for numerical computations.The contribution of evanescent waves (\ref{eq5})
does not have this disadvantage (different aspects of evanescent waves with
respect to the Casimir force are considered in   \cite{57,59,56,58,60}).

The problems arising in the Casimir physics become more pronounced in the limit of
large separations (high temperatures) satisfying the condition
$T\gg \hbar c/(2ak_B)$. In fact, at room temperature $T=300~$K this limit occurs
at separations exceeding $6~\mu$m. In the case of large separations,  (\ref{eq1})
and (\ref{eq4}), (\ref{eq5}) become much simplified. Thus, in  (\ref{eq1})
the total result is given by the zero-frequency term with $l=0$ along, whereas
all terms with $l\geqslant 1$ are exponentially small
\begin{equation}
P(a,T)=-\frac{k_BT}{2\pi}\!\!
\int_0^{\infty}\!\!\!\!d\kb\skb \sum_{\alpha}
\frac{r_{\alpha}^2(0,\kb)e^{-2a\kb}}{1-r_{\alpha}^2(0,\kb)e^{-2a\kb}}.
\label{eq8}
\end{equation}

The condition of large separations can be identically rewritten in the form
$\hbar\omega_c\ll 2k_BT$, where $\omega_c=c/(2a)$ is the characteristic frequency giving the major contribution to the Casimir pressure. Because of this, in the limiting
case of large separations one can use the inequality $\hbar\omega\ll 2k_BT$ and
replace $\coth(\hbar\omega/2k_BT)$ in  (\ref{eq4}) and  (\ref{eq5}) with the first
expansion term in powers of the small parameter
\begin{equation}
\coth\frac{\hbar\omega}{2k_BT}\approx\frac{2k_BT}{\hbar\omega}.
\label{eq9}
\end{equation}
\noindent
Then,  (\ref{eq4}) and  (\ref{eq5}) take the form
\begin{equation}
P^{\rm prop}(a,T)=\frac{k_BT}{\pi^2}\int_{0}^{\infty}\frac{d\omega}{\omega}
\int_0^{k_0}\!\!d\kb\kb \tilde{q}\sum_{\alpha}{\rm Re}
\frac{r_{\alpha}^2\okb e^{2ia\tilde{q}}}{1-r_{\alpha}^2\okb e^{2ia\tilde{q}}},
\label{eq10}
\end{equation}
\noindent
and
\begin{equation}
P^{\rm evan}(a,T)=-\frac{k_BT}{\pi^2}\int_{0}^{\infty}\frac{d\omega}{\omega}
\int_{k_0}^{\infty}\!\!d\kb\kb q\sum_{\alpha}{\rm Im}
\frac{r_{\alpha}^2\okb e^{-2aq}}{1-r_{\alpha}^2\okb e^{-2aq}}.
\label{eq11}
\end{equation}

The quantity (\ref{eq8}) can be easily calculated using the dielectric permittivities
of the Drude and plasma models
\begin{equation}
\ve_{D}(\omega)=1-\frac{\omega_p^2}{\omega(\omega+i\gamma)},
\quad
\ve_{pl}(\omega)=1-\frac{\omega_p^2}{\omega^2},
\label{eq12}
\end{equation}
\noindent
where $\omega_p$ is the plasma frequency and $\gamma$ is the relaxation parameter.
Substituting  (\ref{eq12}) in  (\ref{eq2}), one obtains the values of the
reflection coefficients at zero Matsubara frequency for  the Drude and plasma models
\begin{equation}
r_{D,{\rm TM}}(0,\kb)=1,\qquad r_{D,{\rm TE}}(0,\kb)=0
\label{eq13}
\end{equation}
\noindent
and
\begin{equation}
r_{pl,{\rm TM}}(0,\kb)=1,{\ \ }
r_{pl,{\rm TE}}(0,\kb)=
\frac{c\kb-\sqrt{c^2\skb+\omega_p^2}}{c\kb+\sqrt{c^2\skb+\omega_p^2}},
\label{eq14}
\end{equation}
\noindent
respectively.

Then, substituting  (\ref{eq13}) and (\ref{eq14}) in  (\ref{eq8}), one
finds the following different results \cite{4}:
\begin{equation}
P_D(a,T)=-\frac{k_BT}{8\pi a^3}\zeta(3), \quad
P_{pl}(a,T)=-\frac{k_BT}{4\pi a^3}\zeta(3),
\label{eq15}
\end{equation}
\noindent
where $\zeta(z)$ is the Riemann zeta function. Note that differences between the
theoretical predictions for the thermal Casimir force using the Drude and plasma
models were discussed long ago in   \cite{61} and \cite{62}, respectively.

Now we consider what is the relative role of the propagating and evanescent waves
in the Casimir pressures (\ref{eq15}). Let us start with the pressures calculated
using the Drude model. For the TE polarization of the electromagnetic field, this
question was solved in   \cite{63,63a} using  (\ref{eq10}) and (\ref{eq11}).
It was shown that
\begin{equation}
P_{D,{\rm TE}}^{\rm prop}(a,T)=-\frac{k_BT}{8\pi a^3}\zeta(3), {\ \ }
P_{D,{\rm TE}}^{\rm evan}(a,T)=\frac{k_BT}{8\pi a^3}\zeta(3).
\label{eq16}
\end{equation}
\noindent
Thus, the contribution of evanescent waves leads to the repulsive Casimir force
and cancels the contribution of propagating waves with the result
$P_{D,{\rm TE}}=0$. As for the TM mode, numerical computations performed for the
evanescent waves using  (\ref{eq11}) result in the attractive Casimir pressure
whose magnitude is negligibly small as compared to $P_{D,{\rm TE}}^{\rm evan}$
in  (\ref{eq16}). Then, using the first equality in  (\ref{eq15}) for the
total pressure, one can conclude that
\begin{equation}
P_{D,{\rm TM}}^{\rm prop}(a,T)=-\frac{k_BT}{8\pi a^3}\zeta(3), \quad
P_{D,{\rm TM}}^{\rm evan}(a,T)=0.
\label{eq17}
\end{equation}

We are coming now to the Casimir pressure calculated using the plasma model.
It is given by the second equality in  (\ref{eq15}). The contributions of
evanescent waves (\ref{eq11}) to this result for both the TM and TE modes
are equal to zero because the dielectric permittivity $\ve_{pl}(\omega)$ in
 (\ref{eq12}) and the reflection coefficients in  (\ref{eq6}) are the
real-valued functions. In regard to the propagating waves, the TM and TE
modes contribute to  (\ref{eq10}) equally because at large separations
the characteristic frequency $\omega_c$ goes to zero and the TM and TE
reflection coefficients (\ref{eq16}) go to 1 and --1, respectively, if the
plasma model is used in calculations.

As a result, taking into account the second equality in  (\ref{eq15}),
one obtains
\begin{equation}
P_{pl,{\rm TE}}^{\rm prop}(a,T)=-\frac{k_BT}{8\pi a^3}\zeta(3), \quad
P_{pl,{\rm TE}}^{\rm evan}(a,T)=0
\label{eq18}
\end{equation}
\noindent
and
\begin{equation}
P_{pl,{\rm TM}}^{\rm prop}(a,T)=-\frac{k_BT}{8\pi a^3}\zeta(3), \quad
P_{pl,{\rm TM}}^{\rm evan}(a,T)=0.
\label{eq19}
\end{equation}

{}From  (\ref{eq16})--(\ref{eq19}) we conclude that at large separations
between metallic plates both the TM and TE modes make equal contributions to
the Casimir pressures determined by the propagating waves irrespective of
whether the Drude or the plasma model is used in calculations. As is seen
from  (\ref{eq17}) and (\ref{eq19}), the TM mode also makes equal
(zeroth) contributions to the Casimir pressures computed using either the
Drude or the plasma model.

Thus, according to  (\ref{eq16}) and (\ref{eq18}), the heart of the problem
is in the TE contribution to the Casimir pressure determined by the low-frequency
evanescent waves. This contribution essentially depends on the model of dielectric
permittivity used. However, as discussed in Section~ 1, the required experimental
data concerning the dielectric response of metal to the evanescent waves are few.

Before proceeding to the proposed experiment which could supply us with these
data, we briefly dwell on the phenomenological spatially nonlocal dielectric
permittivity suggested in   \cite{44,45,46,47}. This permittivity combines
the merits of the Drude model by taking into account the dissipation of
conduction electrons and of the plasma model by bringing the Lifshitz theory
in agreement with experiments on measuring the Casimir force and with the requirements
of thermodynamics.

In the spatially nonlocal case it is convenient to express the reflection coefficients
in terms of the surface impedances $Z_{\alpha}\okb$. Thus, the TE reflection
coefficient of our interest is given by \cite{37,64}
\begin{equation}
r_{\rm TE}\okb=\frac{qZ_{\rm TE}\okb+ik_0}{qZ_{\rm TE}\okb-ik_0},
\label{eq20}
\end{equation}
\noindent
where in the approximation of specular reflection the TE impedance is expressed via
the transverse dielectric permittivity $\ve^{\rm Tr}(\omega,\vk)$ as \cite{37,65}
\begin{equation}
Z_{\rm TE}\okb=\frac{ik_0}{\pi}\int_{-\infty}^{\infty}
\frac{dk_z}{k_0^2\ve^{\rm Tr}(\omega,\vk)-\vk^2}.
\label{eq21}
\end{equation}
\noindent
Here, $k_z$ is the wave vector component perpendicular to the plate and
$\vk^2=\skb+k_z^2$. Recall that in the presence of spatial dispersion the response
of metals to perpendicular and parallel to $\vk$ electric filds is described by
the transverse, $\ve^{\rm Tr}(\omega,\vk)$, and longitudinal, $\ve^{\rm L}(\omega,\vk)$,
dielectric permittivities \cite{55,64}. In doing so the TM impedance and respective
TM reflection coefficient are expressed via both of them.

In   \cite{46}, the following transverse dielectric permittivity was suggested to
describe the response of metals to the low-frequency electromagnetic field
\begin{equation}
\ve^{\rm Tr}(\omega,\vk)=1-\frac{\omega_p^2}{\omega(\omega+i\gamma)}
\left(1+i\frac{v^{\rm Tr}k}{\omega}\right),
\label{eq22}
\end{equation}
\noindent
where $k=|\vk|$, the value of $v^{\rm Tr}=1.5v_{\rm F}$, and $v_{\rm F}$ is the
Fermi velocity. This permittivity brings theoretical predictions of the Lifshitz
theory in good agreement with the experimental data of all high-precision
measurements of the Casimir force. In doing so, the theoretical predictions are
scarcely  affected by the form of $\ve^{\rm L}(\omega,\vk)$.

The distinctive feature of the permittivity (\ref{eq22}) is that for the propagating
waves satisfying a condition $\kb<k_0$ it differs little from the standard Drude
 model in  (\ref{eq12}). This is because for the propagating waves the
 magnitude of an addition to unity in the parentheses of  (\ref{eq22}) becomes
 negligibly small
 \begin{equation}
\frac{v^{\rm Tr}k}{\omega}=\frac{v^{\rm Tr}}{c}\frac{k}{k_0}
=\frac{v^{\rm Tr}}{c}\ll 1.
\label{eq23}
\end{equation}
\noindent
As to extremely evanescent waves for which $\kb \gg k_0$, the permittivity
(\ref{eq22}) may significantly deviate from the Drude model.

It should be noted that  the permittivity (\ref{eq22}) is of entirely phenomenological
character and does not lay claim that it is generally applicable. Specifically,
the imaginary part of $\ve^{\rm Tr}$ is positive under the condition
$k<\gamma/v^{\rm Tr}$.
For good metals, such as Au or Cu, this condition is satisfied for
$a\sim 1/(2k)>v^{\rm Tr}/(2\gamma)$, i.e., at separations exceeding a few tens  of
nanometers (we recall that the high-precision experiments  of
  \cite{11,12,13,14,15,16,17,18,19,20,21,22,23} were performed at much larger
separations).

Below we consider the experimental test for the response of metals in the area of
evanescent waves unrelated to the Casimir effect which may give additional information
regarding the validity of dielectric functions (\ref{eq12}) and (\ref{eq22}) in
this area.

%%%%%%%%%%%%%%%%%%%%%%%%%%%%%%%%%%%%%%%%%%%%%%%%%%%%%%%%%%%%%%%%%%%%%%%%%%
\section{Field of oscillating magnetic dipole near a metallic plate}

We consider the field of an oscillating magnetic dipole situated above a thick
metallic plate and show that it bears a close analogy to the Casimir pressure at
large separation between two metallic plates discussed in the previous section.

%%%%%%%%%%%%%%%%%%%%%%__Fig._1__%%%%%%%%%%%%%%%%%%
\begin{figure}[H]
%\vspace*{-0.8cm}
%\centerline{\hspace*{6.5cm}
\includegraphics[width=10.5cm]{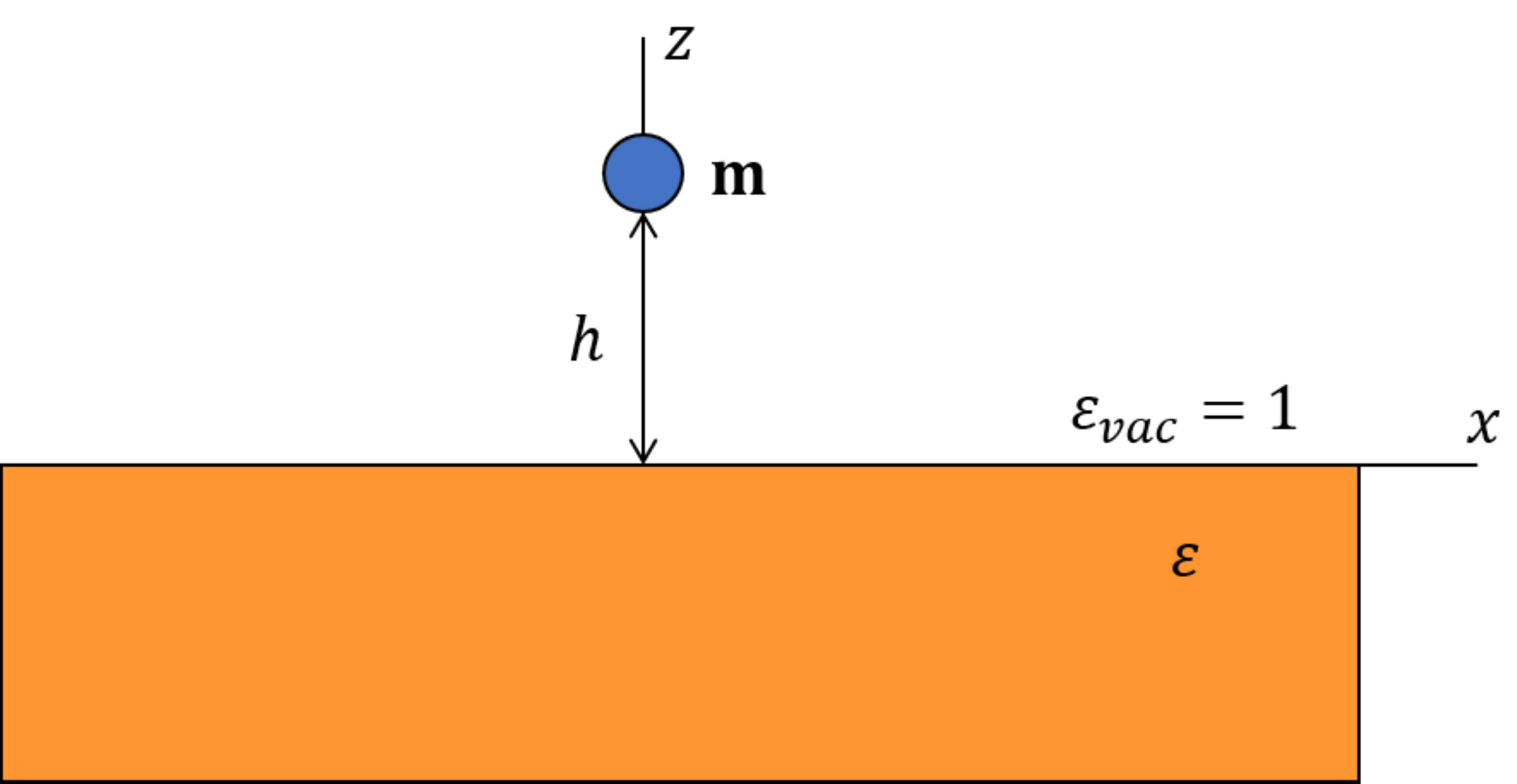}
%\vspace*{-31.5cm}
\caption{\label{fg1}
Magnetic dipole spaced in vacuum above thick metallic plate described
by the dielectric permittivity $\varepsilon$.}
\end{figure}
%%%%%%%%%%%%%%%%%%%%%%%%%%%%%%%%%%%%%%%%%%%%%%%%%%%
Let the magnetic moment of our dipole be directed along the $z$ axis which is
perpendicular to the surface of metallic plate situated in the $(x,y)$ plane
(see Figure~\ref{fg1})
\begin{equation}
\mbox{\boldmath$m$}=(0,\,0,\,m_0e^{-i\omega_{d} t}).
\label{eq24}
\end{equation}
\noindent
The dipole with the oscillation frequency $\omega_{d}$ is spaced in vacuum at the height $h$ above the plate. It is assumed that
$h$ is much larger than the dipole size.

Let us consider first the electromagnetic field of the magnetic dipole (\ref{eq24})
in free space, i.e., in the absence of a conducting plate. Under a condition that
the wavelength $\lambda_d=2\pi c/\omega_d$ is much larger than the dipole size the
solution of this problem is contained in   \cite{66} (\S~72, problem~1).
All fields considered below depend on $t$ as $\exp(-i\omega_d t)$.
The monochromatic  components of the magnetic field
can be represented in the form
\begin{eqnarray}
&&
H_x(\omega_d,\rv)=-m_0\frac{xz}{r^2}\left(\frac{k_d^2}{r}+3i\frac{k_d}{r^2}-
\frac{3}{r^3}\right)e^{ik_dr},
\nonumber\\
&&
H_y(\omega_d,\rv)=-m_0\frac{yz}{r^2}\left(\frac{k_d^2}{r}+3i\frac{k_d}{r^2}-
\frac{3}{r^3}\right)e^{ik_dr},
\label{eq25}\\
&&
H_z(\omega_d,\rv)=m_0\left[\frac{k_d^2}{r}+i\frac{k_d}{r^2}-
\frac{1}{r^3}-\frac{z^2}{r^2}\left(\frac{k_d^2}{r}+3i\frac{k_d}{r^2}-
\frac{3}{r^3}\right)\right]e^{ik_dr},
\nonumber
\end{eqnarray}
\noindent
where the radius-vector $\rv=(x,y,z)$ is directed from the magnetic dipole to
the observation point, $r=|\rv|$, $k_d=\omega_d/c$,
and the Gaussian system of units is used.
In a similar way, the monochromatic components of the electric field
are given by
\begin{equation}
\mbox{\boldmath$E$}(\omega_d,\rv)=im_0k_{d}\left(i\frac{k_d}{r^2}-\frac{1}{r^3}
\right)e^{ik_dr}\left(\begin{array}{rrr}y\\-x\\0\end{array}\right).
\label{eq26}
\end{equation}

Below we consider the quasistationary case of low oscillation frequencies
$\omega_d$ and large $\lambda_d\gg r$. In this case, the magnitudes of components
of the electric
field (\ref{eq26}) are smaller than those of the magnetic field (\ref{eq25})
by the factor of
\begin{equation}
k_dr=\frac{\omega_d r}{c}=\frac{2\pi r}{\lambda_d}\ll 1.
\label{eq27}
\end{equation}
\noindent
Taking into account that in the configuration of suggested experiment
$k_dr\lesssim 10^{-9}$ (see Section~ 5), one can safely neglect by the electric
field as compared to the magnetic one.

Below we also use the Fourier expansion of the field of magnetic dipole (\ref{eq25})
in plane waves $\exp(i\vk_{\bot}\rv_{\bot})$ where $\vk_{\bot}=(k_x,k_y)$ and
$\rv_{\bot}=(x,y)$
\begin{equation}
\mbox{\boldmath$H$}(\omega_d,\rv)=\frac{1}{(2\pi)^2}\int d\vk_{\bot}
e^{i\vk_{\bot}\rv_{\bot}}\mbox{\boldmath$H$}(\omega_d,\vk_{\bot},z).
\label{eq28}
\end{equation}
\noindent
It should be recalled that the third component of the wave vector, $k_z$,
is not independent but is expressed via $k_x$ and $k_y$ by the dispersion
relation $k_z^2\equiv\tilde{q}_d^2=k_d^2-\skb$, see the first equality in  (\ref{eq7}), following from the Maxwell equations.

The explicit expressions for the components of the Fourier transform
$\mbox{\boldmath$H$}(\omega_d,\vk_{\bot},z)$ are (see the Appendix~A where these
expressions are proven)
\begin{eqnarray}
&&
H_{x(y)}(\omega_d,\vk_{\bot},z)=-2\pi im_0k_{x(y)}
{\rm sign}(z)e^{i\tilde{q}_d|z|},
\nonumber \\
&&
H_{z}(\omega_d,\vk_{\bot},z)=2\pi im_0\frac{\skb}{\tilde{q}_d}e^{i\tilde{q}_d|z|}.
\label{eq29}
\end{eqnarray}

Now we find the field of the magnetic dipole in the presence of a thick metallic
plate (see Figure~\ref{fg1}).
We do not specify the explicit form of the reflection coefficients
$r_{\alpha}\dokb$ to take into account both cases of spatially local and
nonlocal response of the plate to the
electromagnetic field.

In fact in our case the reflected wave is fully determined by only the TE
reflection coefficient $\alpha={\rm TE}$. As was shown above, the electric field of
our magnetic dipole is negligibly small in comparison with the magnetic one.
At the same time, the magnetic field of the TM-polarized wave is perpendicular
to the plane of incidence, i.e., parallel to the surface of metallic plate.
Thus, it is not reflected from this plate.

The magnetic field of magnetic dipole in vacuum in the domain above the
metallic plate can be calculated most simply by the method of images \cite{67,68,68a}.
As shown in the Appendix~B, the same result can be obtained using  the
method of the Green tensor.
 To apply the method of images, we use the coordinate system
illustrated in Figure~\ref{fg1} where the magnetic dipole is positioned at the point
$x=y=0$, $z=h$ and the fictitious (image) dipole is in the point
$x=y=0$, $z=-h$. The magnetic moment of the image dipole is in opposition to the
real one and its magnitude depends on the reflectivity properties of the metallic
plate. In the domain above the plate the resulting magnetic field is found as
a superposition of the fields of real and image dipoles. Using  (\ref{eq29}),
we find the Fourier transform of the resulting magnetic field in the presence of
a plate
\begin{eqnarray}
&&
H_{x(y)}^{(p)}(\omega_d,\vk_{\bot},z)=-2\pi im_0k_{x(y)}\left[r_{\rm TE}\dokb
e^{i\tilde{q}_d(z+h)}
+{\rm sign}(z-h)e^{i\tilde{q}_d|z-h|}\right],
\nonumber\\
&&
H_{z}^{(p)}(\omega_d,\vk_{\bot},z)=2\pi im_0\frac{\skb}{\tilde{q}_d}
\left[r_{\rm TE}\dokb
e^{i\tilde{q}_d(z+h)}
+e^{i\tilde{q}_d|z-h|}\right],
\label{eq30}
\end{eqnarray}
\noindent
where the first terms on the right-hand side are related to the image dipole
and the second --- to the real dipole.

The presence of only TE reflection coefficient in  (\ref{eq30}) is explained
by the fact that the TE polarized field has a nonzero component of magnetic
field perpendicular to the surface. Note that for the oscillating electric
dipole the magnetic field is negligibly small, as compared to the electric one,
and, vice versa, the TM polarized field has a nonzero component of the electric
field perpendicular to the surface.

Now we substitute  (\ref{eq30}) into
 (\ref{eq28}) and represent the total magnetic field above the plate in
the form
\begin{equation}
\mbox{\boldmath$H$}^{(p)}(\omega_d,\rv)=\mbox{\boldmath$H$}^{(R)}(\omega_d,\rv)
+\mbox{\boldmath$H$}(\omega_d,\rv).
\label{eq31}
\end{equation}
\noindent
Here, $\mbox{\boldmath$H$}(\omega_d,\rv)$ is the monochromatic field
(\ref{eq25}) of real magnetic dipole where $z$ should be replaced
with $z-h$ and $\mbox{\boldmath$H$}^{(R)}(\omega_d,\rv)$ is the monochromatic
field of image dipole given by
\begin{eqnarray}
&&
H_{x(y)}^{(R)}(\omega_d,\rv)=-\frac{i}{2\pi}m_0\int d\vk_{\bot} \,
r_{\rm TE}\dokb
k_{x(y)}e^{i[\vk_{\bot}\rv_{\bot}+\tilde{q}_d(z+h)]},
\nonumber\\
&&
H_{z}^{(R)}(\omega_d,\rv)=\frac{i}{2\pi}m_0\int d\vk_{\bot} \,r_{\rm TE}\dokb
\frac{\skb}{\tilde{q}_d}e^{i[\vk_{\bot}\rv_{\bot}+\tilde{q}_d(z+h)]}.
\label{eq32}
\end{eqnarray}

The integrals in  (\ref{eq32}) can be calculated  as is described in the
Appendix A for the case of magnetic dipole in the absence of a plate.
Thus, using the polar coordinates defined in (\ref{A2}) and  (\ref{A3}),
the $x$-component, $H_x^{(R)}$, is rewritten as
\begin{equation}
H_{x}^{(R)}(\omega_d,\rv)=-\frac{i}{2\pi}m_0\int_{0}^{\infty}\!\!\! d\kb\skb  \,
r_{\rm TE}\dokb e^{i\tilde{q}_d(z+h)}
\int_{0}^{2\pi}\!\!\!d\psi\cos\psi e^{i\kb\rho\cos(\psi-\varphi)}.
\label{eq33}
\end{equation}
\noindent
Substituting here  (\ref{A5}), one finds
\begin{equation}
H_{x}^{(R)}(\omega_d,\rv)=\frac{m_0x}{\rho}\int_{0}^{\infty}\!\!\! d\kb\skb  \,
J_1(\kb\rho)
r_{\rm TE}\dokb e^{i\tilde{q}_d(z+h)}.
\label{eq34}
\end{equation}

Likewise, the $y$-component in  (\ref{eq32}), $H_y^{(R)}$, is expressed as
\begin{equation}
H_{y}^{(R)}(\omega_d,\rv)=\frac{m_0y}{\rho}\int_{0}^{\infty}\!\!\! d\kb\skb  \,
J_1(\kb\rho)
r_{\rm TE}\dokb e^{i\tilde{q}_d(z+h)}.
\label{eq35}
\end{equation}

Using the polar coordinates, for the $z$-component in  (\ref{eq32}) we obtain
\begin{equation}
H_{z}^{(R)}(\omega_d,\rv)=\frac{im_0}{2\pi}\!\int_{0}^{\infty}\!\!\!d\kb
\frac{k_{\bot}^3}{\tilde{q}_d} \,
r_{\rm TE}\dokb e^{i\tilde{q}_d(z+h)}
\int_{0}^{2\pi}\!\!\!d\psi e^{i\kb\rho\cos(\psi-\varphi)}.
\label{eq36}
\end{equation}
\noindent
Then, applying  (\ref{A12}), we rewrite  (\ref{eq36}) as
\begin{equation}
H_{z}^{(R)}(\omega_d,\rv)=im_0\int_{0}^{\infty}\!\!\!d\kb
\frac{k_{\bot}^3J_0(\kb\rho)}{\tilde{q}_d}
r_{\rm TE}\dokb
e^{i\tilde{q}_d(z+h)}.
\label{eq37}
\end{equation}

Needless to say that both the propagating ($\kb<k_{d}$) and evanescent ($\kb>k_{d}$)
waves contribute to  (\ref{eq34}), (\ref{eq35}), and (\ref{eq37}).
However, in the region $r\ll\lambda_{d}$ under consideration here one can neglect
by the phase in the exponential factors under the integrals. Then, the
contribution of propagating waves in the reflected field, as compared to the
contribution of evanescent ones, is suppressed by the factor
$(k_{d}h)^3=(2\pi h/\lambda_{d})^3$. In the configuration of suggested experiment
$h\sim r$ and $(k_{d}h)^3\sim 10^{-27}$ (see Section~5). This means that the
contribution of propagating waves to  (\ref{eq34}), (\ref{eq35}), and
(\ref{eq37}) does not play any role and one can replace the lower integration
limit in these equations with $k_{d}$. The results of numerical computations
confirm this conclusion.

Finally, we write out the full components of the monochromatic field (\ref{eq31})
of the magnetic dipole   in the presence of metallic plate in the form
convenient for computations
%%%%%%%%%%%%%%%%%%%%%%%%%%%%%%%%%%%%%%%%%%%%
\begin{equation}
H_x^{(p)}(\omega_d,\rv)=\frac{m_0x}{\rho}\!\!\int_{k_d}^{\infty}\!\!\! d\kb\skb
J_1(\kb\rho)r_{\rm TE}\dokb e^{-\sqrt{\skb-k_{d}^2}(z+h)}
\nonumber
\end{equation}
\begin{equation}
-m_0\frac{x(z-h)}{r^2}\left(\frac{k_d^2}{r}+3i\frac{k_d}{r^2}-\frac{3}{r^3}
\right)\,e^{ik_dr}\!\!,
\label{eq38}
\end{equation}
\begin{eqnarray}
&&
H_z^{(p)}(\omega_d,\rv)=m_0\int_{k_d}^{\infty}\!\!\!d\kb
\frac{k_{\bot}^3}{\sqrt{\skb-k_{d}^2}} \,
J_0(\kb\rho)r_{\rm TE}\dokb e^{-\sqrt{\skb-k_{d}^2}(z+h)}
\nonumber\\
&&
+m_0\left[\frac{k_d^2}{r}+i\frac{k_d}{r^2}-\frac{1}{r^3}
-\frac{(z-h)^2}{r^2}\left(\frac{k_d^2}{r}+3i\frac{k_d}{r^2}-\frac{3}{r^3}
\right)\right]\,e^{ik_dr}.
\label{eq39}
\end{eqnarray}

%%%%%%%%%%%%%%%%%%%%%%%%%%%%%%%%%%%%%%%%%%%
\noindent
The expression for $H_y^{(p)}(\omega_d,\rv)$ is similar to  (\ref{eq38})
where $x$ is replaced with $y$.

In fact  (\ref{eq34}),  (\ref{eq35}) and (\ref{eq37}) have much in common with
 (\ref{eq10}) and (\ref{eq11}) presenting the contributions of propagating
and evanescent waves to the Casimir pressure between two parallel metallic
plates at large separations. At first glance  (\ref{eq10}) and (\ref{eq11})
are quite different  from the components of magnetic field
(\ref{eq38}) and (\ref{eq39})
because the former are presented as the integrals over real frequencies
whereas the latter are measured as the function of dipole oscillation
frequency $\omega_d$. Moreover,  (\ref{eq10}) and (\ref{eq11}) contain the
second power of the factor $r_{\alpha}e^{-aq}$ and take into account multiple
reflections on two surfaces whereas  (\ref{eq38}) and (\ref{eq39})
describe the single reflection on one surface.

What is in common, however, is the role of the TE-polarized evanescent waves in both
cases.
The contribution of evanescent waves to the Casimir force at large separations
is given by  (\ref{eq11}). In the case of magnetic field of an oscillating
magnetic dipole located above metallic plate, the contribution of evanescent waves
is given by the first terms in  (\ref{eq38}) and (\ref{eq39}).
By using the dimensionless variables $v=2a\sqrt{\skb-k_{0}^2}$ in  (\ref{eq11}) and
$v=(z+h)\sqrt{\skb-k_d^2}$  in  (\ref{eq38}) and (\ref{eq39}), one can see that
the values of the Casimir force and of the dipole magnetic field are determined
by one and the same integration domain $v\sim 1$.
The lateral components of the field of magnetic dipole in the presence
of metallic plate computed at $z=h$ are determined
by the TE-polarized evanescent waves alone. If we replace the lower integration
limit $k_d$ in   (\ref{eq38}) and (\ref{eq39}) with 0, this does not influence the obtained results, i.e., the propagating waves do not contribute to them.
According to Section~2, the contribution of propagating waves (\ref{eq10})
to the Casimir pressure is common
for both polarizations and for any model of dielectric permittivity used.
The TM-polarized evanescent waves do not contribute to the Casimir pressure irrespective
of the used dielectric function. Thus, again, only the contribution of TE-polarized
evanescent waves eventually determines the value of the total pressure which is
different for different models of the dielectric permittivity. This means that by
measuring magnetic field of the magnetic dipole with fixed oscillation
frequency spaced near a metallic plate one can
test the response of metals to the TE-polarized
evanescent waves in the wide region of real frequencies
contributing to the magnetic field
and simultaneously to the Casimir pressure.
The obtained results can be used to resolve the Casimir puzzle.

%%%%%%%%%%%%%%%%%%%%%%%%%%%%%%%%%%%%%%%%%%%%%%%%%%%%%%%%%%%%%%%%%%%%%%%%%%%%%%%%%%%%
\section{Influence of the response function in the  evanescent domain on the field
of magnetic dipole near metallic plate}

Here, we perform numerical computations of the dipole field (\ref{eq38}) and
(\ref{eq39}) for different oscillator frequencies and different models of the
dielectric permittivity. Taking into account that an impact of metal on the
dipole field is performed through the reflected field in  (\ref{eq31}),
it is advantageous to make all computations at the dipole height $z=h$.
At this height, the lateral component (\ref{eq38}) of the dipole field
coincides with that of the reflected field whereas the relative role of
reflected field in $z$-component (\ref{eq39}) is maximized.

As a metal of the plate, we consider Cu with the Drude parameters
$\omega_p=1.12\times 10^{16}~$rad/s and  $\gamma_p=1.38\times 10^{13}~$rad/s
\cite{72}. The oscillation frequencies of the magnetic dipole should be chosen
so as to gain sufficient reflectance and simultaneously large enough deviation
between theoretical predictions obtained using the Drude and plasma models.

To do the estimate of the corresponding parameters, one can use the  spatially
local reflection coefficients (\ref{eq6}).
We introduce a new variable
\begin{equation}
w=hq_d=h\sqrt{\skb-k_d^2}
\label{eq40}
\end{equation}
\noindent
and rewrite the TE reflection coefficient defined in  (\ref{eq6}) as
\begin{equation}
r_{\rm TE}\dokb=\frac{w-\sqrt{w^2-K(\omega_d)}}{w+\sqrt{w^2-K(\omega_d)}},
\label{eq41}
\end{equation}
\noindent
where
\begin{equation}
K(\omega_d)=[\ve(\omega_d)-1]\frac{\omega_d^2}{\omega_h^2}
\label{eq42}
\end{equation}
\noindent
and $\omega_h=c/h$.

The major contribution to $|K(\omega_d)|$ computed in the region of small frequencies
under consideration here using the dielectric permittivity of the Drude model is
given by its imaginary part
\begin{equation}
|K(\omega_d)|=\frac{\gamma\omega_p^2\omega_d}{(\omega_d^2+\gamma^2)\omega_h^2}
\approx\frac{\omega_p^2\omega_d}{\gamma\omega_h^2}.
\label{eq43}
\end{equation}
\noindent
If $|K(\omega_d)|\gg 1$, the reflections on a metallic surface occur much as in the
case when the plasma model is used.  If $|K(\omega_d)|$ is much less than unity,
 only a small fraction of the magnetic field is reflected from
the surface. Thus, for our purposes, it would be the best to choose
\begin{equation}
|K(\omega_d)|\approx\frac{\omega_p^2\omega_d}{\gamma\omega_h^2}
\leqslant 1,\quad \omega_d\leqslant \Omega\equiv\frac{\gamma\omega_h^2}{\omega_p^2}.
\label{eq44}
\end{equation}
\noindent
For the Drude parameters of Cu indicated above and the typical value of $h=1~$cm
(see Section~ 5), one obtains $\Omega_{\rm Cu}\approx 100~$rad/s. Based on this result,
the computations below are performed for the oscillation frequencies of magnetic
dipole equal to 0.2, 2, 10, 20, and 100~rad/s.

%%%%%%%%%%%%%%%%%%%%%%__Fig._2__%%%%%%%%%%%%%%%%%%
\begin{figure}[H]
\vspace{-9cm}
\centerline{\hspace{-4.5cm}
\includegraphics[width=15cm]{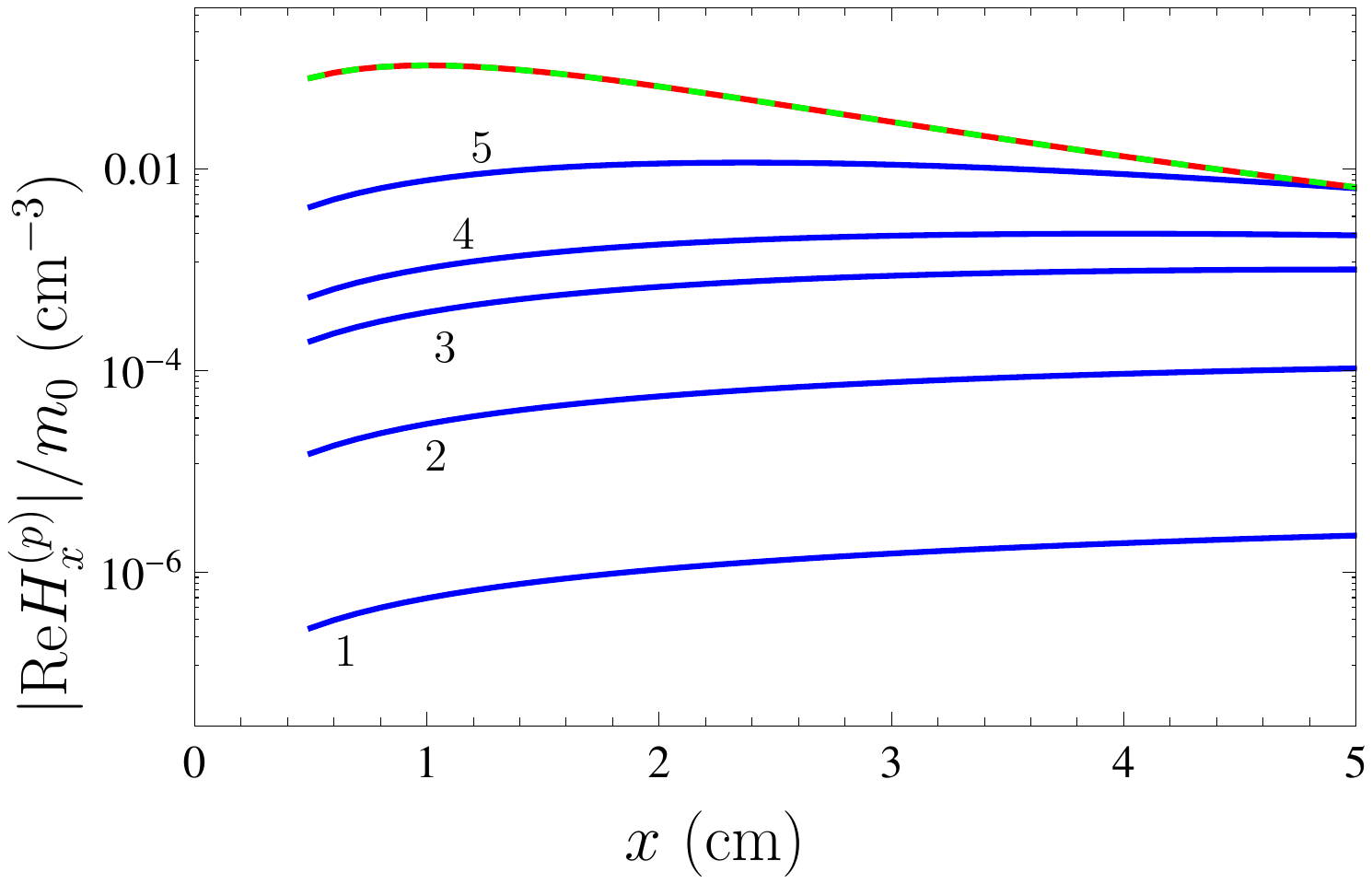}}
\vspace{-4.5cm}
\caption{\label{fg2}
The normalized magnitude of real part of the lateral component
of monochromatic magnetic field reflected from the copper plate
computed using the Drude model as the function of separation
from the magnetic dipole  at the dipole height of 1~cm above
a plate is shown by the solid lines labeled
1, 2, 3, 4, and 5 for the oscillation frequency equal to
0.2, 2, 10, 20, and 100 rad/s, respectively.
The top solid and overlapping with it dashed lines are computed
using the plasma and spatially nonlocal models.}
\end{figure}
%%%%%%%%%%%%%%%%%%%%%%%%%%%%%%%%%%%%%%%%%%%%%%%%%%%
In Figure~\ref{fg2}, we present the computational results for $|{\rm Re}H_x^{(p)}|$
normalized to $m_0$ as a function of separation from the magnetic dipole along the
$x$-axis. In this case, one should put in  (\ref{eq38}) $\rv=(x,0,h)$ and $\rho=x$.
We recall that all computations are performed at $z=h=1~$cm.
Thus, only the reflected field contributes to the result, i.e.,
$H_x^{(p)}$ = $H_x^{(R)}$. The solid lines labeled 1, 2, 3, 4,
and 5 are
computed using the Drude model and the Fresnel reflection coefficient $r_{\rm TE}$
defined in  (\ref{eq6}) for the dipole oscillation frequencies
equal to 0.2, 2, 10, 20, and 100~rad/s, respectively. The top solid line is computed
using the plasma model and the coefficient $r_{\rm TE}$
defined in  (\ref{eq6}). Under the condition (\ref{eq27}) it does not depend
on the value of $\omega$. The top dashed line overlapping with the solid one is
computed using the spatially nonlocal dielectric permittivity (\ref{eq22}) and
the impedance reflection coefficient (\ref{eq20}). The Fermi velocity
$v_{\rm F}^{\rm Cu}=1.6\times 10^6~$m/s was determined under an assumption of the
spherical Fermi surface. The top dashed line is also independent on the value
of $\omega_d$ as well as in
the case of the top solid line plotted for the plasma model.

{}From Figure~\ref{fg2}, one can see that, as expected, the difference between the
values of $|{\rm Re}H_x^{(p)}|$ computed using the Drude and plasma models increases
with decreasing oscillation frequency of the magnetic dipole. This difference reaches
several orders of magnitude and can be observed experimentally (see Section~5).

\begin{figure}[H]
\vspace{-9cm}
\centerline{\hspace{-4.5cm}
\includegraphics[width=15cm]{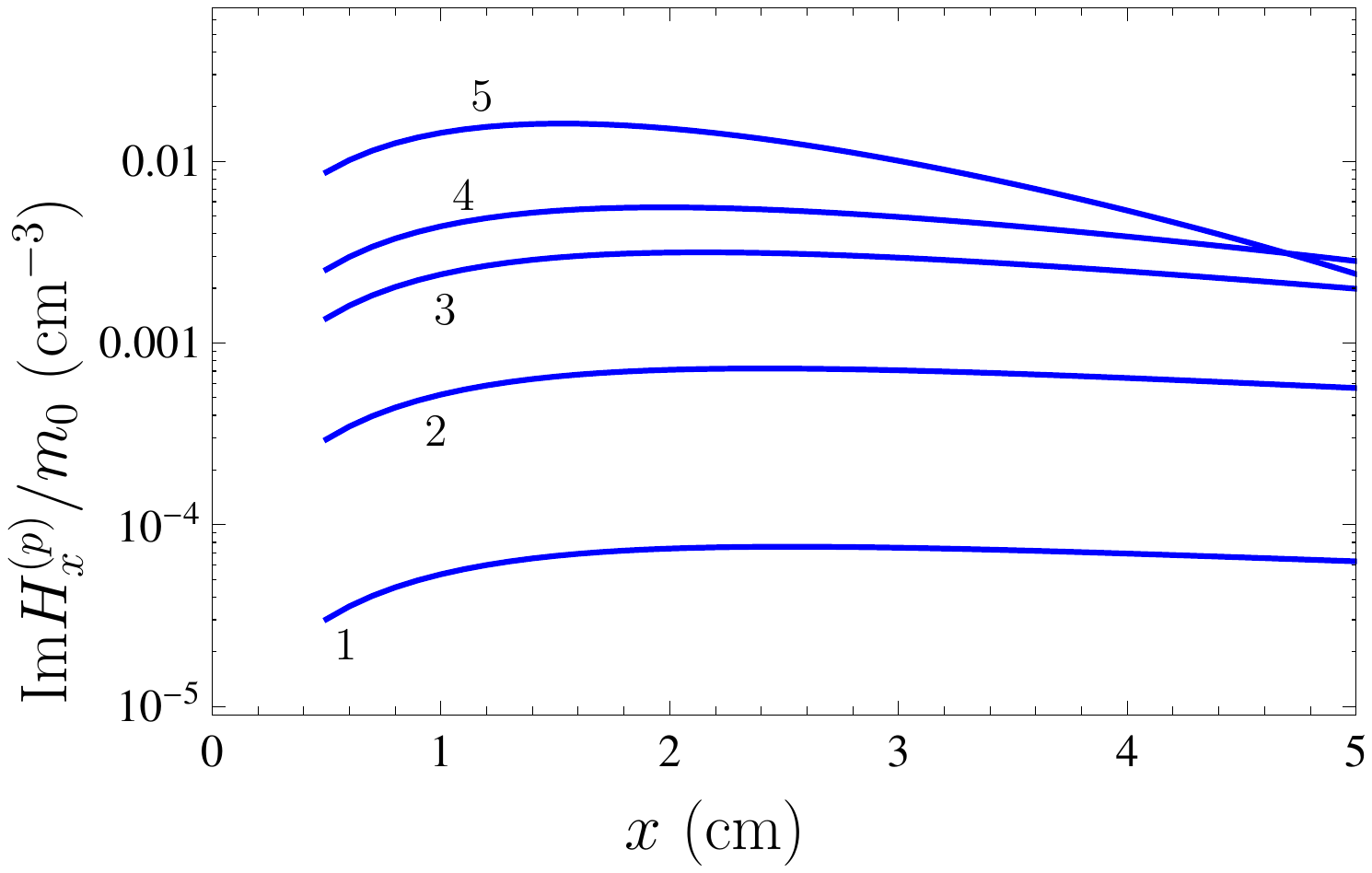}}
\vspace{-4.5cm}
\caption{\label{fg3}
The normalized imaginary part of the lateral component
of monochromatic magnetic field reflected from the copper plate
computed using the Drude model as the function of separation
from the magnetic dipole at the dipole height of 1~cm above
a plate is shown by the solid lines labeled
1, 2, 3, 4, and 5 for the oscillation frequency equal to
0.2, 2, 10, 20, and 100 rad/s, respectively.}
\end{figure}

Similar results for  the imaginary part of $H_x^{(p)}$ = $H_x^{(R)}$ normalized to $m_0$ are shown
in Figure~\ref{fg3}. Here, again, the  lines labeled 1, 2, 3, 4, and 5 are
computed by  (\ref{eq38})
using the Drude model  for the dipole oscillation frequencies
 of 0.2, 2, 10, 20, and 100~rad/s, respectively.
 For the plasma model, the reflection coefficient $r_{\rm TE}$ in  (\ref{eq38})
 is real so that the obtained result is ${\rm Im}H_x^{(p)}/m_0=0$.
 If the spatially nonlocal dielectric permittivity defined in  (\ref{eq22}) is
 used in computations, the obtained values of ${\rm Im}H_x^{(p)}/m_0$ are by ten
 orders of magnitudes smaller than those shown by the line labeled 1 in Figure~\ref{fg3}.
%%%%%%%%%%%%%%%%%%%%%%__Fig._3__%%%%%%%%%%%%%%%%%%
%%%%%%%%%%%%%%%%%%%%%%%%%%%%%%%%%%%%%%%%%%%%%%%%%%%

By comparing Figure~\ref{fg3} with Figure~\ref{fg2}, it is seen that   ${\rm Im}H_x^{(p)}$
is much greater than  $|{\rm Re}H_x^{(p)}|$ for each value of oscillation frequency
if the Drude model is used in computations. If, however, the plasma or spatially
nonlocal models are used, the magnitude of the real part, $|{\rm Re}H_x^{(p)}|$,
is far in excess of ${\rm Im}H_x^{(p)}$ which is either equal to zero or negligibly
small.

%%%%%%%%%%%%%%%%%%%%%__Fig._4__%%%%%%%%%%%%%%%%%%
\begin{figure}[ptb]
\vspace{-4cm}
\centerline{\hspace{-5cm}
\includegraphics[width=15cm]{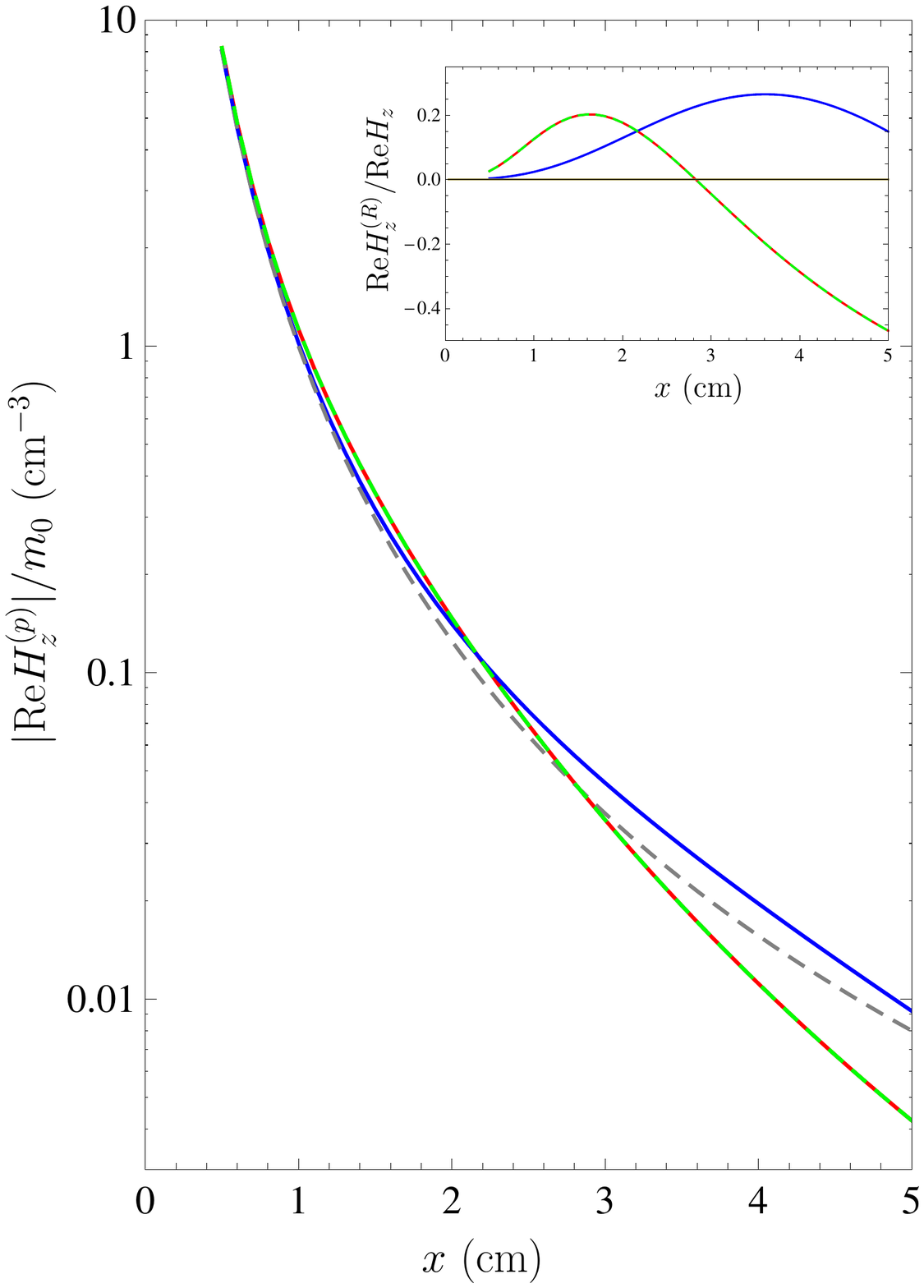}}
\vspace{-3cm}
\caption{\label{fg4}
The normalized magnitudes of real part of the z-component
of monochromatic magnetic field above the copper plate
computed using the Drude model and by the plasma and
spatially nonlocal models at the dipole height of 1~cm
 for the oscillation
frequency of 100 rad/s as the functions of separation
from the magnetic dipole are shown by the solid line
and by another solid line overlapping with the short-dashed
one, respectively. The long-dashed line demonstrates
the normalized magnitude of real part of the z-component
of magnetic field of oscillating dipole in the absence
of metallic plate. The inset shows the ratio of real part
of the z-component of reflected field to the same quantity
for the dipole field in the absence of metallic plate
computed using the Drude model (the solid line) and the
plasma or spatially nonlocal models (the solid line
overlapping with the short-dashed one).}
\end{figure}
%%%%%%%%%%%%%%%%%%%%%%%%%%%%%%%%%%%%%%%%%%%%%%%%%%%
 For completeness we also consider the component $H_z^{(p)}$ given in  (\ref{eq39})
 which contains both contributions of the source dipole field and its part reflected
 from the plate. The computational results for $|{\rm Re}H_z^{(p)}|$ at $z=h=1~$cm
 normalized to $m_0$ as the function of separation from the magnetic dipole along
 the $x$-axis are presented in Figure~\ref{fg4}.
The solid line is obtained for $\omega_d=100~$rad/s using the Drude model and another
solid line overlapping with the short-dashed one, which does not depend on frequency,
is found by means of the plasma model and the spatially nonlocal model (\ref{eq22}).
For comparison purposes, the long-dashed line shows the quantity  $|{\rm Re}H_z|/m_0$
computed in the absence of metallic plate for a magnetic dipole in free space.
It is also frequency-independent for the frequency region under consideration.

{}From Figure~\ref{fg4} it is seen that at separations exceeding 2~cm there is a
considerable deviation between the
theoretical predictions of the Drude model, on the one
hand, and the plasma and our spatially nonlocal models, on the other hand.
This deviation takes the largest value at $x=5~$cm. It is seen also that at large
separations the impact of metallic plate on theoretical predictions obtained using the
Drude model is far less than can be obtained when the plasma or spatially nonlocal
models are used. Because of this,
the measurement of the $z$-component of magnetic field can
be considered as complementary to measurement of the lateral component
(according to Figure~\ref{fg2} for $\omega_d=100~$rad/s the largest deviation between
different theoretical predictions for the lateral component occurs at $x=1~$cm).

In the inset to Figure~\ref{fg4}, we also plot the ratio of ${\rm Re}H_z^{(R)}$ to
${\rm Re}H_z$ as the function of separation where the first quantity characterizes
the reflected field and the second one is for the dipole field in free space in the
absence of metallic plate. The solid line is computed using the Drude model and
another solid line overlapping with the dashed one is found by means of the plasma
and spatially nonlocal models. According to the inset, the quantity
${\rm Re}H_z^{(R)}$ determined by the reflected field changes its sign with
separation if it is computed using the plasma model and the  spatially nonlocal model
(\ref{eq22}) but preserves its sign if the Drude model is used in computations.
Some computational results for ${\rm Im}H_z$ are presented in Section~5.

%%%%%%%%%%%%%%%%%%%%%%__Fig._5__%%%%%%%%%%%%%%%%%%
\begin{figure}[ptb]
\vspace{-4cm}
\centerline{\hspace{-5cm}
\includegraphics[width=15cm]{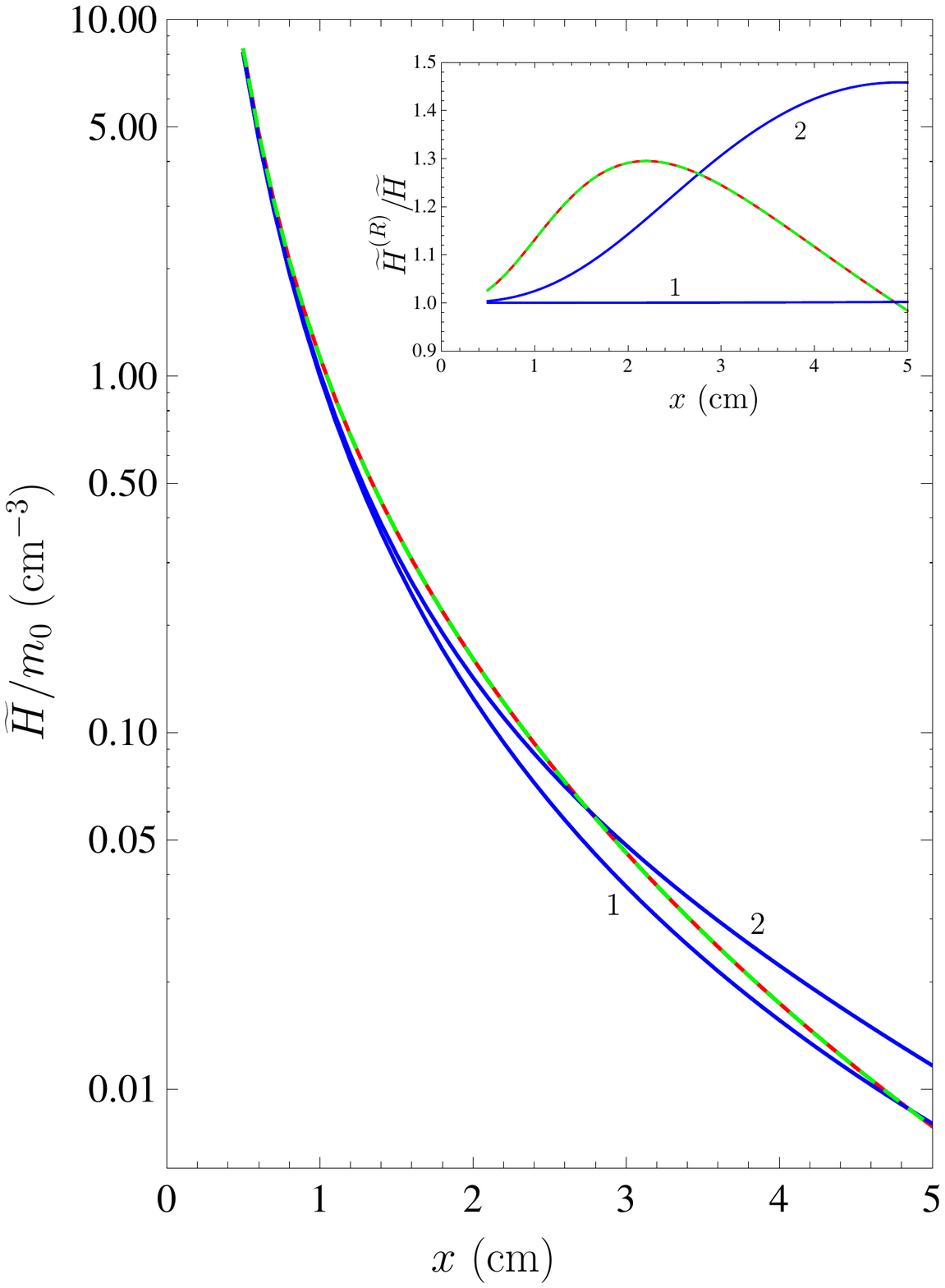}}
\vspace{-3cm}
\caption{\label{fg5}
The normalized magnitude of total
monochromatic magnetic field above the copper plate
computed using the Drude model at the dipole height of 1~cm
for the oscillation
frequency of 0.2 and 100 rad/s as the function of separation
from the magnetic dipole is shown by the solid lines 1 and 2,
respectively. Another solid line overlapping with the dashed
one shows the same quantity computed using the plasma and
spatially nonlocal models. The inset shows the ratio of
magnitudes of the reflected and total field computed using
the Drude model (lines 1 and 2 for the dipole oscillation
frequencies of 0.2 and 100 rad/s, respectively) and the
plasma and spatially nonlocal models (the solid line
overlapping with the dashed one).   }
\end{figure}
%%%%%%%%%%%%%%%%%%%%%%%%%%%%%%%%%%%%%%%%%%%%%%%%%%%
It is interesting also to illustrate the dependence on separation of the total field
magnitude
\begin{equation}
\widetilde{\!H}(\omega_d,\rv)=\sqrt{{|H_x^{(p)}(\omega_d,\rv)|}^2+
{|H_z^{(p)}(\omega_d,\rv)|}^2}.
\label{eq45}
\end{equation}
\noindent
In Figure~\ref{fg5}, the quantity $\widetilde{\!H}(\omega,r)$ normalized to $m_0$ is
shown as the function of separation from the magnetic dipole by the solid lines
labeled 1 and 2, which are computed using the Drude model for the dipole oscillation
frequencies of 0.2 and 100~rad/s, respectively. Another solid line overlapping
with the dashed one is computed employing the plasma and the spatially nonlocal
models. These lines do not depend on the value of oscillation frequency.

{}From Figure~\ref{fg5}, it is seen that at separations of 2.67 and 4.85~cm the
theoretical predictions of all three models coincide for the dipole oscillation
frequency equal to 100 and 0.2~rad/s, respectively. At these separations, there are,
however, significant deviations between the theoretical predictions of the Drude mode,
on the one hand, and the plasma and spatially nonlocal models, on the other hand,
if the oscillation frequencies of 0.2 and 100~rad/s are used, respectively.

In the inset to Figure~\ref{fg5}, we plot as a function of separation the ratio of
the reflected,
\begin{equation}
\widetilde{\!H}\vphantom{H}^{(R)}(\omega_d,\rv)=
\sqrt{{|H_x^{(R)}(\omega_d,\rv)|}^2+{|H_z^{(R)}(\omega_d,\rv)|}^2},
\label{eq46}
\end{equation}
\noindent
to the total (\ref{eq45}) magnitudes of the field.
The solid lines 1 and 2 are computed using
the Drude model for the dipole oscillation frequencies of 0.2 and 100~rad/s, respectively,
whereas another solid line overlapping with the dashed one using the plasma and the
spatially nonlocal models. As is seen from the inset, for the oscillation
frequency of 0.2~rad/s the quantity
$\widetilde{\!H}\vphantom{H}^{(R)}/\widetilde{\!H}$ is
separation-independent. The largest difference between the theoretical predictions
is reached at 5~cm for the dipole oscillation frequency of 100~rad/s.

The above computations are for a plate made of copper. The obtained results are,
however, applicable to any metallic material with appropriately chosen frequencies
of the magnetic dipole. Thus, for the plate made of B-doped Si in the metallic state
used in the experiment of   \cite{73} it holds
$\omega_p^{\rm Si}=7.0\times 10^{14}~$rad/s and
$\gamma_p^{\rm Si}=1.5\times 10^{14}~$rad/s.
Then, using  (\ref{eq44}), one obtains $\Omega_{\rm Si}=2.75\times 10^{5}~$rad/s.
Because of this, by choosing the dipole oscillation frequencies in the same
proportion to $\Omega_{\rm Si}$ as above to $\Omega_{\rm Cu}$, i.e.,
$\Omega_{\rm Si}/500$, $\Omega_{\rm Si}/50$, $\Omega_{\rm Si}/10$,
$\Omega_{\rm Si}/5$, and $\Omega_{\rm Si}$, one arrives to the same computational
results as were obtained for a copper plate. This means that Figures~\ref{fg2}--\ref{fg5}
plotted for the case of Cu plate are also valid for a B-doped Si plate if the
oscillation frequencies of the magnetic dipole 0.2, 2, 10, 20, and 100~rad/s are replaced
with $5.5\times 10^2$, $5.5\times 10^3$, $2.75\times 10^4$, $5.5\times 10^4$,
$2.75\times 10^5~$rad/s, respectively.

%%%%%%%%%%%%%%%%%%%%%%%%%%%%%%%%%%%%%%%%%%%%%%%%%%%%%%%%%%%%%%%%%%%%%%%%%%%%%%%%%%%
\section{{\bf\textit {Experimentum crucis}} for the Casimir puzzle}

The oscillating magnetic dipole with a magnetic moment (\ref{eq24}) is generated
by an alternating current $I_0\exp(-i\omega_d t)$ through a circular turn or a small
coil. In this case
\begin{equation}
m_0=\frac{1}{c}\pi NI_0R^2,
\label{eq47}
\end{equation}
\noindent
where $R$ is the turn radius and $N$ is the number of turns. In accordance to
Figure~\ref{fg1}, the loop should be located in the plane $z=h$ with a center at the
point $(0,0,h)$. The surface of metallic plate is in the plane $z=0$.
As shown in Section~3, if the magnetic field is measured at the points with
coordinates $(x,0,h)$, i.e., along the line parallel to the $x$-axis at the
height $h$, its lateral component is completely determined by reflections from
the surface of a metal [see  (\ref{eq38})].

For a qualitative estimation of the coil parameters, we can consider the case
of total reflection, $r_{\rm TE}=-1$, which is close to metal plate
described by the plasma model. Then the Fourier transform of the
$x$-component of the magnetic field is
obtained from   (\ref{eq30}) where the second term does not contribute because $z=h$.
As a result, under the condition  (\ref{eq27}), the $x$-component of the field is given by the last
term on the right-hand side of  (\ref{eq25}) where one should replace $z$ with
$z+h=2h$
\begin{equation}
H_x^{(p)}(x)=\frac{6m_0xh}{(x^2+4h^2)^{5/2}}=\frac{m_0}{h^3}
\,\frac{6\tilde{x}}{(4+\tilde{x}^2)^{5/2}}.
\label{eq48}
\end{equation}
\noindent
Here, $\tilde{x}=x/h$.

The scale factor $m_0/h^3$ in  (\ref{eq48}) determines the field magnitude.
Thus, in order to have the greater field, one has to use larger $m_0$, i.e.,
the larger current, turn radius, and the number of turns, but smaller distance
$h$ between the magnetic dipole and the metallic plate. These conditions are
somewhat contradictory.
The calculations made in Section~3 assume that both $h$ and the distance $x$ between
the source and the observation point are large in comparison  with
 the dipole size.
This means that in order to decrease $h$ one should
decrease the dipole size accordingly. For too small coils, however, it would
be difficult to ensure large currents and the number of turns.

In the literature, one can find examples of small coils manufactured by the
methods of mini- \cite{74} and micro- \cite{75,76} technologies
(micro-electromagnets for atom manipulation have long been created \cite{77}).
The coils of such kind contain of about 10 turns with a radius of 1~mm and
have height  of about 1~mm.
These coils are able to support the current
up to $I_0=3\times 10^9~$statA (this is
equal to 1~A in SI). The respective magnetic dipole moment computed by
 (\ref{eq47}) is equal to
\begin{equation}
m_0=3.14\times 10^{-2}~\mbox{erg/Oe}
=3.14\times 10^{-5}~\mbox{A\,m}^2.
\label{eq49}
\end{equation}

For estimation of the produced lateral field, we put  $x=h=10~$mm so that $\tilde{x}=1$.
Then  (\ref{eq48}) leads to
\begin{equation}
H_x^{(p)}=3.36\times 10^{-3}\,\mbox{Oe}= 3.36\times 10^{-7}\,\mbox{T}=
0.27\,\mbox{A/m}.
\label{eq49a}
\end{equation}
\noindent
By using the value of $m_0$ from  (\ref{eq49}), one can
see that the obtained field is in a very good agreement with the top line in
Figure~\ref{fg2} at $x=1~$cm which was computed using the plasma model and the spatially
nonlocal model of   \cite{46}.

Thus, the numerical data in Figures~   \ref{fg2}--\ref{fg5} in combination with the
value of $m_0$ in  (\ref{eq49}) can be used for calculation of the magnetic field
produced by the oscillating magnetic dipole at different separations and oscillation
frequencies depending on the used model of electromagnetic response of metal to the
low-frequency evanescent waves. As an example, if the Drude model is used and the
 dipole oscillation frequency is equal to 100~rad/s (see the line 5 in Figure~\ref{fg2}),
one has
\begin{equation}
|{\rm Re}H_x^{(p)}|\lesssim
3.36\times 10^{-4}\,\,\mbox{Oe}= 3.36\times 10^{-8}\,\,\mbox{T}=
0.027\,\mbox{A/m},
\label{eq50}
\end{equation}
\noindent
i.e., by more than an order of magnitude smaller field than for the plasma model
in the separation region from 1 to 2~cm. In this case, the parameter $k_0r$ defined
in  (\ref{eq27}) is of the order of $10^{-9}$. The Drude model predicts all the
more damping of the lateral magnetic field for smaller oscillation frequencies
(see the lines 1--4 in Figure~\ref{fg2}).

The obtained values of the magnetic field are quite measurable. Using different
methods and laboratory techniques, the presently available limit for the magnetic
field resolution is down to $10^{-13}~$T \cite{78,79,80}.
This resolution is already reached \cite{80a,80b} by means of traditional search coil
magnetometers consisting of several micro induction coils \cite{80c,80d}.
The same sensitivity is provided by the SQUID magnetometers \cite{80e,80f,80g}.
Some novel types of magnetometers using the magnetoresistance and
magnetoelectric effects, as well as the optically pumped atomic magnetometers,
possess even higher resolution when measuring alternating magnetic fields (see \cite{79,80h} for
a review). In the proposed experiment one should measure the magnetic fields of the
order of tens of nT which are by several orders of magnitude larger than the
resolution limit of currently available magnetic sensors. Therefore, the
systematic effects, i.e., from the industrial or thermal noise, although deserve
careful attention, should not be a crucial problem.
Based on this, measuring
the magnetic field of an oscillating magnetic dipole can be considered as the
{\it experimentum crucis} which can provide valuable information concerning the
electromagnetic response of metals to the evanescent waves of low frequency.
Depending on the magnitude of measured magnetic field, one will have the possibility
to clearly discriminate between the description of metal in the range of evanescent
waves by means of the Drude model, as the first option, or by means of the plasma
model or spatially nonlocal model as the second option.

Direct demonstration of an inadequacy of the Drude model in the range of low-frequency
evanescent waves by means of an experiment performed in the area of classical
electrodynamic phenomena would be of great help to resolution of the Casimir
puzzle discussed in Sections~1 and 2. The Casimir effect is the quantum phenomenon
dealing with the concept of quantum vacuum.
As indicated above, high precision experiments of
  \cite{11,12,13,14,15,16,17,18,19,20,21,22,23}
on measuring the Casimir force have already demonstrated that the
theoretical predictions of the Lifshitz theory obtained using the
Drude model at separations exceeding 150~nm are excluded by the
measurement data. An independent confirmation of a conclusion
that the Drude model does not work in the area of evanescent waves
would also be of prime importance for several divisions of optics
and condensed matter physics dealing with evanescent waves.

Keeping in mind subsequent experiments, in addition to the
computational results presented in Section 4, here we report the
real and imaginary parts of $H_x^{(p)}$ and $H_z^{(p)}$ computed as the
functions of dipole frequency for the above parameters of the
experimental configuration [i.e., for the value of $m_0$
defined in  (\ref{eq49}), $z=h=1$~cm, $y=0$] with different
response functions of a metal and for several values of the
separation $x$ between the dipole and the point of observation.
The computations are performed using  (\ref{eq38}) and
(\ref{eq39}) for a copper plate.

\begin{figure}[ptb]
\vspace{-5cm}
\centerline{\hspace{-2.5cm}
\includegraphics[width=25cm]{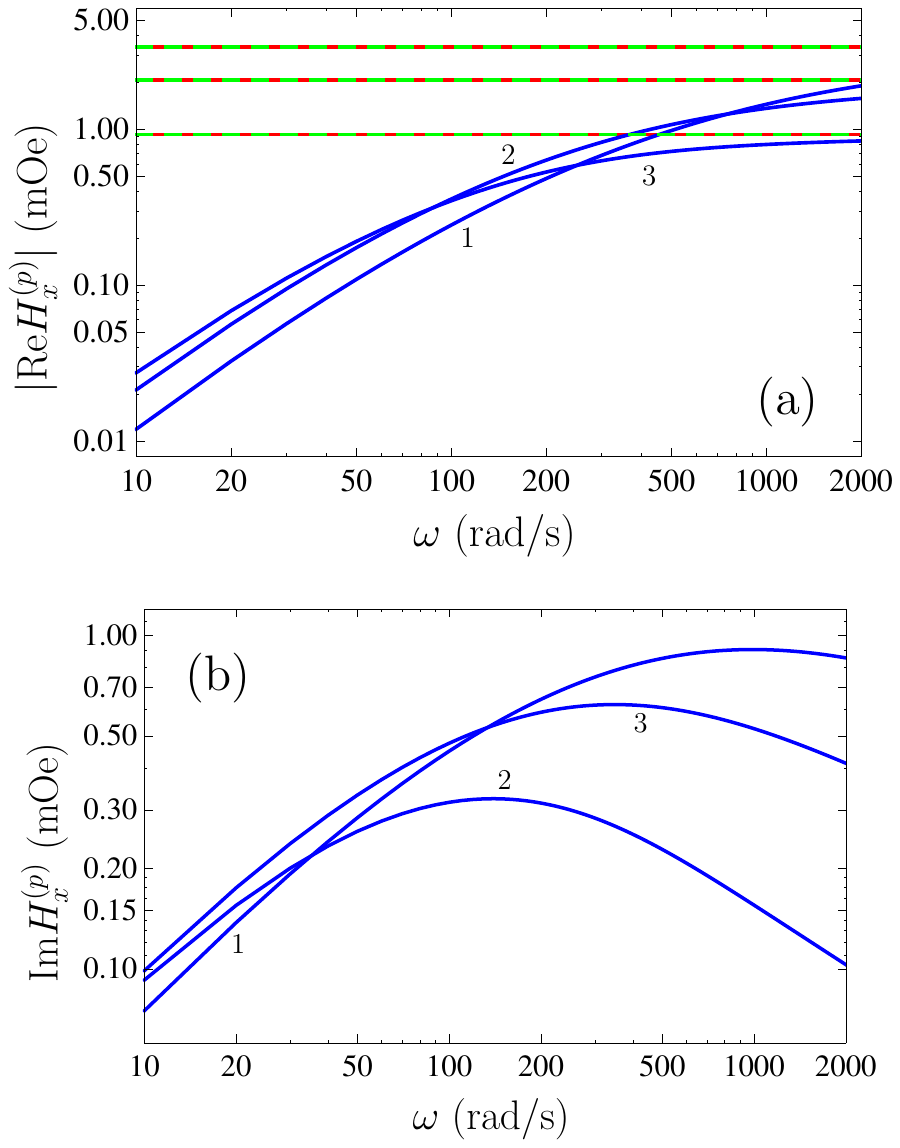}}
\vspace{-16cm}
\caption{\label{fg6}
The magnitude of real part (a) and the imaginary part (b) of
the lateral component of monochromatic magnetic field reflected from
the copper plate computed using the Drude model at the dipole
height of 1~cm above
a plate is shown by the
solid lines labeled 1, 2, and 3 as the function of oscillation
frequency at separations from the magnetic dipole equal to 10, 20,
and 30 mm, respectively. The three solid lines overlapping with the
dashed ones counted from top to bottom (a) are computed using
the plasma and spatially nonlocal models at separations of 10,
20, and 30 nm, respectively.}
\end{figure}

In Figure~\ref{fg6}(a), we plot the computational results for
$|{\rm Re}H_x^{(p)}|$ = $|{\rm Re}H_x^{(R)}|$ as a function of dipole
oscillation frequency. The
solid lines labeled 1, 2, and 3 are computed by using the Drude
model at separations from the magnetic dipole equal to $x$=10, 20,
and 30~mm, respectively. The computational results obtained using
the plasma model and the spatially nonlocal model (\ref{eq22})
at 10, 20, and 30~mm are shown by the three solid lines and
overlapping with them dashed lines counted from the top
to  bottom, respectively. In Figure~\ref{fg6}(b), similar
computational results obtained using the Drude model for
$|{\rm Im}H_x^{(p)}|$ = $|{\rm Im}H_x^{(R)}|$ are presented with the
same notations. If the
plasma model is used in computations, one arrives to ${\rm Im}H_x^{(p)}=0$.
As to the spatially nonlocal dielectric permittivity (\ref{eq22}),
it gives ${\rm Im}H_x^{(p)}$ that is smaller
by the factor of $10^{-8}$
 comparing to those shown in Figure~\ref{fg6}(b).

%%%%%%%%%%%%%%%%%%%%%%__Fig._6__%%%%%%%%%%%%%%%%%%
%%%%%%%%%%%%%%%%%%%%%%%%%%%%%%%%%%%%%%%%%%%%%%%%%%%
As one can see from Figure~\ref{fg6}(a),
 different theoretical predictions for $|{\rm Re}H_x^{(p)}|$
can differ significantly. For  $x=10$~mm and $\omega_d=10$~rad/s
the field computed using the plasma model is 280 times larger
than that for the  Drude model.
For the Drude model  the magnitude of ${\rm Re}H_x^{(p)}$ increases
with the frequency but stays constant for the plasma model.
Although the difference between predictions of these models is
reduced, for $\omega_d=100$~rad/s
the predictions still differ by a factor of 14 which is
 sufficient for the experimental discrimination. As to
Figure~\ref{fg6}(b), the values of ${\rm Im}H_x^{(p)}$ shown by any line
computed using the Drude model are much outside the theoretical
predictions based on the plasma and spatially nonlocal
models where ${\rm Im}H_x^{(p)}$vanishes or nearly vanishes.

%%%%%%%%%%%%%%%%%%%%%%__Fig._7__%%%%%%%%%%%%%%%%%%
\begin{figure}[ptb]
\vspace{-5cm}
\centerline{\hspace{-2.5cm}
\includegraphics[width=25cm]{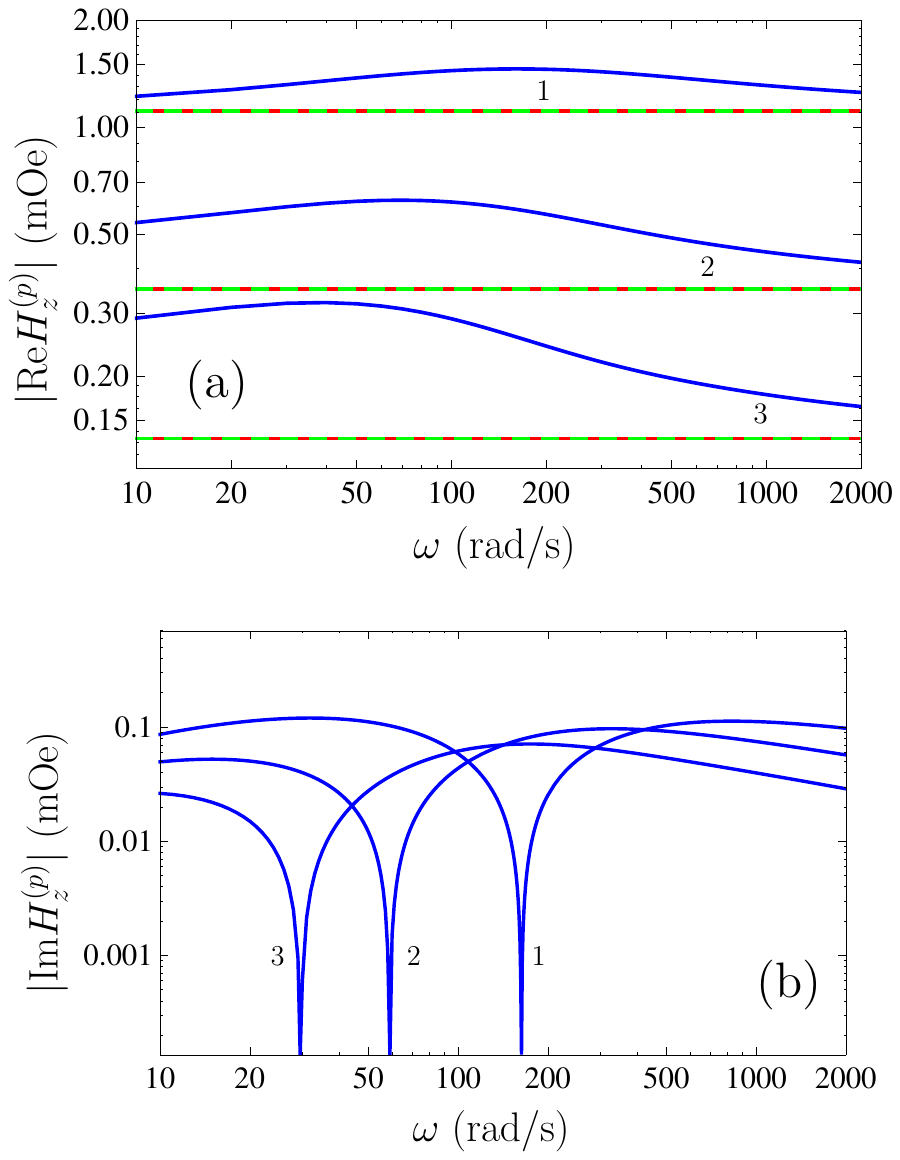}}
\vspace{-16cm}
\caption{\label{fg7}
The magnitudes of real part (a) and the imaginary part (b) of
the normal component of monochromatic magnetic field in the domain
above the copper plate computed using the Drude model at the dipole
height of 1~cm  are shown by
the solid lines labeled 1, 2, and 3 as the functions of oscillation
frequency at separations from the magnetic dipole equal to 30, 40,
and 50 mm, respectively. The three solid lines labeled 1, 2, and 3
overlapping with the dashed ones (a) are computed using
the plasma and spatially nonlocal models at separations of 30,
40, and 50 nm, respectively. }
\end{figure}
%%%%%%%%%%%%%%%%%%%%%%%%%%%%%%%%%%%%%%%%%%%%%%%%%%%
In Figure~\ref{fg7}(a), using the same values of all experimental
parameters, we plot the computational results for $|{\rm Re}H_z^{(p)}|$
as a function of the oscillation frequency of the magnetic
dipole. The results obtained using the Drude and plasma (or the
spatially nonlocal) models are shown by the three pairs of solid
and solid-dashed lines, respectively. These pairs are labeled
as 1, 2, and 3 for separations from the magnetic dipole equal to
30, 40, and 50~mm, respectively. The choice of separations, which
is different from that in Figure~\ref{fg6}, is caused by the
fact that at larger separations one obtains greater differences
between the $z$-components of magnetic field computed using
different models of the dielectric permittivity (see Figure~\ref{fg4}).
The computational results for $|{\rm Im}H_z^{(p)}|$ obtained using the
Drude model are shown in Figure~\ref{fg7}(b) with the same notations.
In common with above, the plasma model leads to ${\rm Im}H_z^{(p)}=0$ and
the magnitudes of ${\rm Im}H_z^{(p)}$ computed using the spatially nonlocal
model (\ref{eq22}) are much below those shown in Figure~\ref{fg7}(b).

As is seen in Figure~\ref{fg7}(a), the theoretical predictions for
$|{\rm Re}H_z^{(p)}|$ obtained using the Drude and plasma models are
not so much different as for $|{\rm Re}H_x^{(p)}|$. This is caused by
a contribution to the $z$-component of the dipole field in the
absence of metallic plate. Thus, measuring the lateral component
of the magnetic field is preferable for a clear discrimination
between different models of the dielectric permittivity.
It should be noted, however, that the imaginary part of the field can be
determined by simultaneous measurements of the field magnitude and the
phase shift. In doing so, if some nonzero phase shift relative
to the magnetic moment is registered, this would be in favor of the
Drude model. {}From
Figure~\ref{fg7}(b) one can also see that ${\rm Im}H_z^{(p)}$
computed using the Drude model possesses an interesting property
by changing its sign from positive to negative at the dipole
oscillation frequencies equal to 29.4, 58.9, and 162.8~rad/s when
the measurements are performed at separations from the dipole of
50, 40, and 30~mm, respectively.

%%%%%%%%%%%%%%%%%%%%%%%%%%%%%%%%%%%%%%%%%%%%%%%%%%%%%%%%%%%%%%%%%%%%%%%%%%%%%%%%%%%

\section{Discussion}

In the foregoing, we have demonstrated that the long-standing problem
of a disagreement between the theoretical predictions of fundamental
Lifshitz theory and numerous high-precision experiments on measuring
the Casimir force at separations exceeding 150 nm is determined by
the role of TE-polarized low-frequency evanescent waves. The analysis
of literature sources shows that the electromagnetic response of metals
to the evanescent waves of this type has not yet been sufficiently
explored experimentally.
Taking into account that the predictions of the Lifshitz
theory come into conflict with the measurement data when the
low-frequency electromagnetic response of metals is described by
the well-established Drude model taking into account the relaxation
properties of conduction electrons, it is very desirable to
independently verify whether or not this model is applicable in
the range of low-frequency evanescent waves.

In this paper, we have derived an expression for the electromagnetic
field of an oscillating magnetic dipole spaced above a thick metallic
plate. According to our results obtained under the reasonable
experimental conditions, the lateral components of
magnetic field of this dipole are
entirely determined by the TE-polarized evanescent waves whereas
the electric field turns out to be negligibly small. The
derivation was performed both by the method of images and using
the Green tensor of the boundary problem in the general case of
a spatially nonlocal electromagnetic response of the plate metal
with the coinciding results.

The real and imaginary parts of the components of magnetic field
above the copper plate were computed by describing the
electromagnetic response of copper using the dissipative Drude
model, dissipationless plasma model and the spatially nonlocal
phenomenological dielectric permittivity of \cite{46} which
takes into account the relaxation properties of conduction electrons
and simultaneously brings the theoretical predictions of the
Lifshitz theory in agreement with all high-precision experiments
on measuring the Casimir force. It was shown that for the typical
parameters of the magnetic dipole designed in the form of 1-mm
coil the lateral components of magnetic field computed
using the Drude model, as the first option, and the plasma model
or the spatially nonlocal model of \cite{46}, as the second
option,  differ by up to several orders of magnitude depending
on the oscillation frequency of the dipole. In doing so, the
magnitudes of the field component remain quite measurable using
the available laboratory techniques. Based on this, we have
proposed an experiment which can independently either validate
or disprove an applicability of the Drude model in the range
of low-frequency TE-polarized evanescent waves.

The suggested experiment would elucidate the roots of the
Casimir puzzle by making clear whether a disagreement between
the theoretical predictions and the measurement data is caused
by an inadequacy of the Drude model in the range of evanescent
waves or by some other problems.
Because of this, it could be called the {\it experimentum
crucis}. Based on the studies of the Casimir force in graphene
systems, one could argue in favor of the inadequacy of the Drude
model. The point is that the spatially nonlocal response of
graphene was derived on the basis of first principles of
quantum electrodynamics via the formalism of the polarization
tensor \cite{81,82}.
As mentioned in Section~1, using the respective nonlocal dielectric
functions, the theoretical predictions of the Lifshitz theory
were shown to be in perfect agreement with the requirements of
thermodynamics \cite{83,83a} and with experiments on measuring the
Casimir force in graphene systems \cite{84,85}.

\section{Conclusions}
According to our results, the suggested experiment will be
capable to clearly discriminate between the magnetic fields
computed using the Drude and plasma models but not between
the fields computed using the plasma model and the spatially
nonlocal permittivity of \cite{46}. The latter, in
any case, does not claim a complete description of the
electromagnetic response of metals but can be considered
as an example that through the spatial nonlocality in the
framework of the Lifshitz theory it is possible to reach an
agreement with the requirements of thermodynamics and the
measurement data of high-precision experiments having regard
to the relaxation properties of conduction electrons (see
also \cite{86} which considers other nonlocal
permittivities with the same aim).

To conclude, the fundamental theoretical description of the
response of metals to both the propagating and evanescent
waves could be derived in future using the methods of quantum
field theory \cite{87} like it was already done for graphene.
The first steps in this direction were made in \cite{88}.
In the meantime, the results of the proposed {\it experimentum
crucis} may not only shed light for the Casimir puzzle, but
also find applications in a number of topics in optics and
condensed matter physics which deal with the evanescent waves
including the studies on surface plasmon polaritons, near-field
optical microscopy, total internal reflection and many others.
%%%%%%%%%%%%%%%%%%%%%%%%%%%%%%%%%%%%%%%%%%%%%%%%%%%%%%%%%%%%%%%%%%%%%%%%%%%%%%

%%%%%%%%%%%%%%%%%%%%%%%%%%%%%%%%%%%%
\authorcontributions{
Conceptualization, Galina L. Klimchitskaya, Vladimir M. Mostepanenko and Vitaly B. Svetovoy;
Investigation, Galina L. Klimchitskaya, Vladimir M. Mostepanenko and Vitaly B. Svetovoy;
Writing – original draft,
Vladimir M. Mostepanenko; Writing – review \& editing, Galina L. Klimchitskaya
and Vitaly B. Svetovoy.}

%%%%%%%%%%%%%%%%%%%%%%%%%%%%%%%%%%%%%%%%%%
\funding{
G.~L.~K.\ and
V.~M.~M.\ were partially supported by the Peter the Great Saint
Petersburg Polytechnic University in the framework of the Russian state
assignment for basic research (Project No.\ FSEG-2020-0024). The work
of V.~M.~M.\ was also supported by the Kazan Federal University
Strategic Academic Leadership Program. V.~B.~S.~was partially supported
by the Russian Science Foundation, grant No.\ 20-19-00214.

}

%%%%%%%%%%%%%%%%%%%%%%%%%%%%%%%%%%%%%%%%%%
%\conflictsofinterest{The authors declare no conflict of interest.}

%%%%%%%%%%%%%%%%%%%%%%%%%%%%%%%%%%%%%%%%%%%%%%
\appendixtitles{yes}
\appendixstart
\appendix
\section{Fourier transform of the field of magnetic dipole }
\setcounter{equation}{0}
\renewcommand{\theequation}{A\arabic{equation}}

Here, we present the proof of  (\ref{eq29}). For this purpose we substitute the
first equality in  (\ref{eq29}) into the first component of the Fourier
transform (\ref{eq28}) and obtain
\begin{equation}
H_x(\omega_d,\rv)=-\frac{i}{2\pi}{\rm sign}(z)\int d\vk_{\bot}k_x
e^{i\vk_{\bot}\rv_{\bot}}\,e^{i\tilde{q}_d|z|}.
\label{A1}
\end{equation}
\noindent
Introducing the polar coordinates
\begin{equation}
\vk_{\bot}=(k_x,k_y)=(\kb\cos\psi,\kb\sin\psi),
\qquad
\rv_{\bot}=(x,y)=(\rho\cos\varphi,\rho\sin\varphi),
\label{A2}
\end{equation}
\noindent
one finds
\begin{equation}
\vk_{\bot}\rv_{\bot}=k_xx+k_yy=\kb\rho\cos(\psi-\varphi).
\label{A3}
\end{equation}

Then, using  (\ref{A3}), one can rewrite  (\ref{A1}) in the form
\begin{equation}
H_x(\omega_d,\rv)=-\frac{i}{2\pi}m_0 {\rm sign}(z)\int_0^{\infty}\!\!\!d\kb \skb
e^{i\tilde{q}_d|z|}
\int_0^{2\pi}\!\!\!d\psi\cos\psi e^{i\kb\rho\cos(\psi-\varphi)}.
\label{A4}
\end{equation}

The integral with respect to $\psi$ in  (\ref{A4}) can be found using the
integral representation for the Bessel functions $J_n(x)$ (see   \cite{69})
\begin{equation}
\int_0^{2\pi}\!\!\!d\psi\cos\psi
e^{i\kb\rho\cos(\psi-\varphi)}=2\pi i J_1(\kb\rho)
\cos\varphi.
\label{A5}
\end{equation}
\noindent
Then  (\ref{A4}) takes the form
\begin{equation}
H_x(\omega_d,\rv)=m_0{\rm sign}(z)\cos\varphi\int_0^{\infty}\!\!\!d\kb \skb
J_1(\kb\rho)e^{i\tilde{q}_d|z|}.
\label{A6}
\end{equation}

The integral in  (\ref{A6}) is calculated by using the following result which
is derived from   \cite{70} (equation 2.12.10.10) by differentiation with
respect to the parameter $p$:
\begin{equation}
\int_0^{\infty}\!\!d\kb k_{\bot}^{\nu+1} e^{-|z|\sqrt{\skb+\tau^2}}
J_{\nu}(\kb\rho)
=\sqrt{\frac{2}{\pi}}\rho^{\nu}|z|\tau^{\frac{3}{2}+\nu}
(z^2+\rho^2)^{-\frac{\nu}{2}-\frac{3}{4}}K_{\nu+\frac{3}{2}}(\tau\sqrt{z^2+\rho^2}),
\label{A7}
\end{equation}
\noindent
where $K_{\nu}(x)$ is the Bessel function of imaginary argument.
Note that  (\ref{A7}) does not coincide with the expression 2.12.20.6 given
in   \cite{70} just for this integral. The point is that the expression 2.12.20.6
is in contradiction with the correct expression 2.12.20.20 and contains two typos:
the power of $\tau$ on the right-hand side, which is indicated as 3/2, must be replaced
with $3/2+\nu$ and the power of $(z^2+\rho^2)$, which is shown as $-\nu-3/4$,
must be replaced with $-\nu/2-3/4$.

Our integral in  (\ref{A6}) is obtained from  (\ref{A7}) as a particular case
$\nu=1$ and $\tau=-ik_0$
\begin{equation}
\int_0^{\infty}\!\!\!d\kb \skb J_1(\kb\rho)e^{-|z|\sqrt{\skb-k_d^2}}
=\sqrt{\frac{2}{\pi}}\rho|z|\sqrt{-i}k_d^{\frac{5}{2}}
(z^2+\rho^2)^{-\frac{5}{4}}
K_{5/2}(-ik_d\sqrt{z^2+\rho^2}).
\label{A8}
\end{equation}
\noindent
Notice that although  (\ref{A7}) was derived under a condition  ${\rm Re}\,\tau>0$,
it can be analytically continued to the point ${\rm Re}\,\tau=0$ leading to
 (\ref{A8}).

The Bessel function of imaginary argument in   (\ref{A8}) can be presented in
terms of the elementary functions \cite{70}
\begin{equation}
K_{5/2}(x)=\sqrt{\frac{\pi}{2}}\frac{x^2+3x+3}{x^{5/2}}\,e^{-x}.
\label{A9}
\end{equation}

Substituting  (\ref{A9}) taken for $x=-ik_0r$ (we recall that $z^2+\rho^2=r^2$)
in  (\ref{A8}), one obtains
\begin{equation}
\int_0^{\infty}\!\!\!d\kb \skb J_1(\kb\rho)e^{-|z|\sqrt{\skb-k_d^2}}
=\rho|z|\frac{3-3ik_dr-k_d^2r^2}{r^5}\,e^{ik_dr}.
\label{A10}
\end{equation}

Finally, after the substitution of (\ref{A10}) in  (\ref{A6}), we arrive to the sought for
first equality in  (\ref{eq25}). The second expression in  (\ref{eq29}) for
the Fourier component $H_y(\omega_d,\vk_{\bot},z)$ is proven in perfect analogy to the
above with a replacement of $k_x$ with $k_y$.

Now we deal with the third expression in   (\ref{eq29}) for
the Fourier component $H_z(\omega_d,\vk_{\bot},z)$. Substituting
it in the $z$ component
of  (\ref{eq28}) and using the polar coordinates (\ref{A2}) and  (\ref{A3})
one obtains
\begin{equation}
H_z(\omega_d,\rv)=\frac{im_0}{2\pi}\!
\int_0^{\infty}\!\!\!d\kb\frac{k_{\bot}^3}{\tilde{q}_d}
e^{i\tilde{q}_d|z|}
\int_0^{2\pi}\!\!\!d\psi e^{i\kb\rho\cos(\psi-\varphi)}.
\label{A11}
\end{equation}

Using the result 2.5.27.19. in   \cite{71}
\begin{equation}
\int_0^{2\pi}\!\!\!d\psi e^{i\kb\rho\cos(\psi-\varphi)}=
2\pi J_0(\kb\rho),
\label{A12}
\end{equation}
\noindent
Equation (\ref{A11}) can be rewritten in the form
\begin{equation}
H_z(\omega_d,\rv)=m_0\!\int_0^{\infty}\!\!\! d\kb
\frac{k_{\bot}^3 J_0(\kb\rho)}{\sqrt{\skb-k_d^2}}
e^{-|z|\sqrt{\skb-k_d^2}}.
\label{A13}
\end{equation}

With the help of equality \cite{70}
\begin{equation}
J_0(\kb\rho)=\frac{2}{\kb\rho}J_1(\kb\rho)-J_2(\kb\rho)
\label{A14}
\end{equation}
\noindent
Equation (\ref{A13}) can be transformed to
\begin{equation}
H_z(\omega_d,\rv)=m_0\left[\frac{2}{\rho}\int_0^{\infty}\!
\!\!\!d\kb\frac{k_{\bot}^2 J_1(\kb\rho)}{\sqrt{\skb-k_d^2}}
e^{-|z|\sqrt{\skb-k_d^2}}-\int_0^{\infty}\!\!\!d\kb
\frac{k_{\bot}^3 J_2(\kb\rho)}{\sqrt{\skb-k_d^2}}
e^{-|z|\sqrt{\skb-k_d^2}}\right].
\label{A15}
\end{equation}
\noindent
Both integrals in (\ref{A15}) are calculated using the result 2.12.10.10 in
  \cite{70} with $\nu=1$ and $\nu=2$ which is analytically continued to the
point ${\rm Re}\,\tau={\rm Re}\,(-ik_d)=0$. Then   (\ref{A15}) is rewritten as
\begin{eqnarray}
&&
H_z(\omega_d,\rv)=m_0\sqrt{\frac{2}{\pi}}\sqrt{i}k_d^{3/2}(z^2+\rho^2)^{-3/4}
\left[2K_{3/2}(-ik_d\sqrt{z^2+\rho^2})\right.
\nonumber\\
&&~~~~~~~~~
-\left.i\rho^2k_d(z^2+\rho^2)^{-1/2}K_{5/2}(-ik_d\sqrt{z^2+\rho^2})\right].
\label{A16}
\end{eqnarray}
Taking into account  (\ref{A9}) and similar identity \cite{70}
\begin{equation}
K_{3/2}(x)=\sqrt{\frac{\pi}{2}}\frac{x+1}{x^{3/2}}\,e^{-x}
\label{A17}
\end{equation}
at $x=-ik_dr=-ik_d\sqrt{z^2+\rho^2}$, one arrives from  (\ref{A16}) to
the third expression in  (\ref{eq25}).

This concludes the proof of  (\ref{eq29}).

%%%%%%%%%%%%%%%%%%%%%%%%%%%%%%%%%%%%%%%%
%%__Appendix__B__%%%%
\section{Field derivation by the Green function method }
\setcounter{equation}{0}
\renewcommand{\theequation}{B\arabic{equation}}

Lifshitz and co-workers \cite{6,6a,7} have used the Green function method to deduce the famous formula for the Casimir force. In this Appendix, we use this method to find the field of a magnetic dipole above a conductive plane derived in the main text
[see  (\ref{eq30})] by the method of images.
We do this in the general case of a spatially nonlocal response
of the plane material. In accordance with the main text,
we are interested only in TE (s) polarization.

In the Green function method, an elementary current is the source of the field. To reproduce the field of magnetic dipole, this elementary current can be chosen as
(along with the main text, the Gaussian system of units is used)
\begin{equation}\label{eq:current}
 \mbox{\boldmath$J$}(\rv)=-c \mbox{\boldmath$m$} \times\mathbf{\nabla}\delta(\rv).
\end{equation}
\noindent
By applying the Biot-Savart law, it can be easily  checked that this current
reproduces the field of magnetic dipole in empty space
[see  (\ref{eq25}) and (\ref{eq26})].

We are looking for the electromagnetic field above the conductive plane at $z=0$
when the source is located in the point $\rv_0=(0, 0, h)$. If the magnetic
moment $\mbox{\boldmath$m$}=\mbox{\boldmath$m$}_0 e^{-i\omega_d t}$ oscillates
with the frequency
$\omega_d$, then all the fields oscillate with the same frequency and the
corresponding monochromatic components obey the Maxwell equations
\begin{equation}\label{eq:Max_E}
  \nabla\times\mbox{\boldmath$E$}(\omega_d,\rv)=ik_d\mbox{\boldmath$H$}(\omega_d,\rv),
\end{equation}
\begin{equation}\label{eq:Max_H}
   \nabla\times\mbox{\boldmath$H$}(\omega_d,\rv)=
   \frac{4\pi}{c}\mbox{\boldmath$J$}(\omega_d,\rv)-ik_d\mbox{\boldmath$E$}(\omega_d,\rv),
\end{equation}
where $k_d=\omega_d/c$ is the absolute value of the wave vector. In vacuum the magnetic fields {\boldmath$B$} and {\boldmath$H$} coincide in the Gaussian system. The electric field is related to the Green tensor $G_{ij}(\omega_d,\rv,\rv^{\prime})$ as
\begin{equation}\label{eq:E_Green}
  E_i(\omega_d,\rv)=\int d\rv^{\prime}
  G_{ij}(\omega_d,\rv,\rv^{\prime})J_j(\omega_d,\rv^{\prime}),
\end{equation}
where the Latin indices $i,\,j=1,\,2,\,3$
enumerate the components of the fields and tensors and a summation is made
over the repeated indices. Excluding {\boldmath$H$} from  (\ref{eq:Max_E})
and (\ref{eq:Max_H}), and expressing {\boldmath$E$} via the Green tensor
with  (\ref{eq:E_Green}), one finds the equation for $G_{il}$:
\begin{equation}\label{eq:Green}
\partial_i (\partial_j G_{jl})-(\nabla^2+k_d^2) G_{il}=i\frac{4\pi k_d}{c} \delta_{il}\delta(\rv-\rv^{\prime}),
\end{equation}
where the derivatives are taken on the argument $\rv$.
Similar equation has been used to derive the Casimir force in   \cite{7}.

Since in the $(x,y)$-plane the problem is homogeneous, it is useful to separate
the lateral components and denote them by Greek indices $\alpha, \beta, \ldots=1,\,2$,
so that $x_1=x$, $x_2=y$, $k_1=k_x$, and $k_2=k_y$.
The homogeneity also means that the Green tensor depends only on the difference $x_\alpha-x'_\alpha$ and one can make the Fourier expansion in these coordinates
\begin{equation}\label{eq:Fourier}
  G_{ij}(\omega_d,\rv,\rv^{\prime})=\int\!\!
  \frac{d \vk_{\bot}}{(2\pi)^2}e^{ik_{\alpha} (x_\alpha-x'_\alpha)}
  G_{ij}(\omega_d,k_\perp,z,z'),
\end{equation}
where $\vk_{\bot}=(k_1,k_2)$.
Applying this expansion to  (\ref{eq:Green}), one finds a closed-form equation for the lateral components of the tensor
\begin{equation}
  \left(\frac{\partial^2}{\partial z^2}+\tilde{q}_d^2\right)
  \left(\delta_{\alpha\gamma}+ \frac{k_{\alpha} k_{\gamma}}{\tilde{q}_d^2}\right) G_{\gamma\beta}(\omega_d,k_\perp,z,z')
=
  -i\frac{4\pi k_d}{c}\delta_{\alpha\beta}\delta(z-z'),
  \label{eq:G_lat}
\end{equation}
where $\tilde{q}_d=\sqrt{k_d^2-\skb}$. The solution of this equation is
\begin{equation}\label{eq:G_via_D}
  G_{\alpha\beta}=\left(\delta_{\alpha\beta}-
  \frac{k_\alpha k_\beta}{k_d^2}\right)G(\omega_d,k_\perp,z,z'),
\end{equation}
where the function $G(\omega_d,k_\perp,z,z')$ is the solution of the equation
\begin{equation}\label{eq:G_problem}
   \left(\frac{\partial^2}{\partial z^2}+\tilde{q}_d^2\right)
   G(\omega_d,k_\perp,z,z')=-i\frac{4\pi k_d}{c}\delta(z-z')
\end{equation}
with appropriate boundary conditions.

If the magnetic moment of the source is directed along the $z$-axis
$\mbox{\boldmath$m$}=(0,0,m_0)$, then this source generates only TE (s) polarization.
For this case it is sufficient to know only the lateral components of the Green
tensor. Using  (\ref{eq:current}), (\ref{eq:E_Green}), and (\ref{eq:G_via_D}),
one finds the electric field components in the presence of metallic plate
\begin{equation}\label{eq:E_field}
E_\alpha^{(p)}(\omega_d,\vk_{\bot},z) =ic m_0 \epsilon_{\alpha\beta}k_\beta G(\omega_d,k_\perp,z,h), \quad E_z =0,
\end{equation}
where $\epsilon_{\alpha\beta}$ is the 2D antisymmetric tensor
(Levi-Civita symbol). Similarly, using  (\ref{eq:Max_E})
the magnetic field is
\begin{equation}
  H_\alpha^{(p)}(\omega_d,\vk_{\bot},z)=c m_0\frac{k_\alpha}{k_d}
   \frac{\partial}{\partial z}G(\omega_d,k_\perp,z,h),
\nonumber
\end{equation}
\begin{equation}
H_z^{(p)}(\omega_d,\vk_{\bot},z)=-i c m_0\frac{\skb}{k_d}
G(\omega_d,k_\perp,z,h).
\label{eq:H_field}
\end{equation}

Thus, it is desirable to solve the boundary problem for  (\ref{eq:G_problem}).
The first boundary condition is $G\to 0$ for  $z\to\infty$.
Furthermore,
at the surface of the reflecting plane the lateral components of the electric
and magnetic fields have to be continuous.
However, such a boundary condition is very inconvenient for a spatially
nonlocal material response. Since we are interested only in the field above
the plane, it is more constructive to use the surface impedance boundary
conditions, when there is no need to know the field in the material.
For the TE (s) polarization the impedance is supposed to be a known
function $Z_{\rm TE}(\omega_d,k_\perp)$ that is defined as the ratio
\begin{equation}\label{eq:Z_impedance}
  Z_{\rm TE}(\omega_d,k_\perp)=
  -\frac{E_\alpha (\omega_d,\vk_{\bot}0)}{\epsilon_{\alpha\beta}
  H_\beta(\omega_d,\vk_{\bot}0)}.
\end{equation}
\noindent
Using here the expressions (\ref{eq:E_field}) and (\ref{eq:H_field}),
one finds the second boundary condition
\begin{equation}\label{eq:second_bc}
 ik_dG(\omega_d,k_\perp,0,h) +
 Z_{\rm TE}(\omega_d,k_\perp) \left.\frac{\partial}{\partial z}
 G(\omega_d,k_\perp,z,h)\right|_{z=0}\!\!\!=0.
\end{equation}
\noindent
The solution of the boundary problem for  (\ref{eq:G_problem}) is the following:
\begin{equation}\label{eq:G_solution}
  G(\omega_d,k_\perp,z,h)=\frac{2\pi k_d}{c \tilde{q}_d}
 \left [r_{\rm TE} e^{i\tilde{q}_d(z+h)}-e^{i\tilde{q}_d|z-h|}\right],
\end{equation}
\noindent
where $r_{\rm TE}$ is the reflection coefficient expressed via the impedance of the material (\ref{eq:Z_impedance}) by  (\ref{eq20}).

Substituting the solution (\ref{eq:G_solution}) in  (\ref{eq:H_field}),
we reproduce the expressions (\ref{eq30}).

%%%%%%%%%%%%%%%%%%%%%%%%%%%%%%%%%%%%%%%%%%
\end{paracol}
\reftitle{References}

%%%%%%%%%%%%%%%%%%%%%%%%%%%%%%%%%%%%%%%%%%
\end{document}